\documentclass{aa}
\usepackage[utf8]{inputenc}
\bibpunct{(}{)}{;}{a}{}{,} 
\usepackage{color}

\title{Synthetic simulations of the extragalactic sky seen by \emph{eROSITA}}
\subtitle{I. Pre-launch selection functions from Monte-Carlo simulations}

\author{N.~Clerc \inst{1,3} \and M.~E.~Ramos-Ceja \inst{2} \and J.~Ridl\inst{1} \and G.~Lamer\inst{4} \and H.~Brunner\inst{1} \and F.~Hofmann \inst{1} \and J.~Comparat \inst{1} \and F.~Pacaud\inst{2} \and F.~K{\"a}fer\inst{1} \and T.~H.~Reiprich\inst{2} \and A.~Merloni\inst{1} \and C.~Schmid\inst{5} \and T.~Brand\inst{5}  \and J.~Wilms\inst{5} \and P.~Friedrich\inst{1} \and A.~Finoguenov\inst{1,6} \and T.~Dauser\inst{5} \and I.~Kreykenbohm\inst{5}}

\institute{Max-Planck Institut f{\"u}r extraterrestrische Physik, Postfach 1312, 85741 Garching bei M{\"u}nchen, Germany
    \and
    Argelander-Institut f{\"u}r Astronomie, Universit{\"a}t Bonn, Auf dem H{\"u}gel 71, 53121 Bonn, Germany 
    \and
     IRAP, Université de Toulouse, CNRS, UPS, CNES, Toulouse, France\\ \email{nicolas.clerc@irap.omp.eu}
     \and 
     Leibniz-Institut f{\"u}r Astrophysik Potsdam (AIP), An der Sternwarte 16, D-14482 Potsdam, Germany
     \and
     Dr. Karl Remeis-Observatory and ECAP, Sternwartstr. 7, 96049 Bamberg, Germany
     \and
    Department of Physics, University of Helsinki, Gustaf H{\"a}llstr{\"o}min katu 2a, FI-00014 Helsinki, Finland} 
    
\date{October 2017}

\abstract{
    Studies of galaxy clusters provide stringent constraints on models of structure formation. Provided that selection effects are under control, large X-ray surveys are well suited to derive cosmological parameters, in particular those governing the dark energy equation of state.
}{
    We forecast the capabilities of the all-sky eROSITA (the extended ROentgen Survey with an Imaging Telescope Array) survey to be achieved by the early 2020s. We bring special attention to modeling the entire chain from photon emission to source detection and cataloguing.
}{
    The selection function of galaxy clusters for the upcoming eROSITA mission is investigated by means of extensive and dedicated Monte-Carlo simulations. Employing a combination of accurate instrument characterization and of state-of-the-art source detection technique, we determine a cluster detection efficiency based on the cluster fluxes and sizes.
}{
    Using this eROSITA cluster selection function, we find that eROSITA will detect a total of $\sim 10^5$ clusters in the extra-galactic sky. This number of clusters will allow eROSITA to put stringent constraints on cosmological models. We show that incomplete assumptions on selection effects, such as neglecting the distribution of cluster sizes, induce a bias in the derived value of cosmological parameters.
}{
    Synthetic simulations of the eROSITA sky capture the essential characteristics impacting the next-generation galaxy cluster surveys and they highlight parameters requiring tight monitoring in order to avoid biases in cosmological analyses.
}

\begin{document}

\maketitle

\section{Introduction}

Clusters of galaxies are the most massive matter halos. They formed last in the history of the Universe by a hierarchical growth of structures in the Hubble expansion flow. Their presence, observed space density and mass distributions confirm the standard cosmological model \citep[e.g.][]{hasselfield2013, mantz2014, planck2016, deHaan2016}, making galaxy clusters powerful probes of cosmological parameters, such as the dark energy content and its equation of state \citep[e.g.][]{vikhlinin2009}, see also \citet{allen2011} for a review. The different components of galaxy clusters (dark matter halo, intra-cluster medium, galaxies and relativistic particles) allow using different observational techniques to identify and to study them. Among such techniques, X-ray observations stand out, since clusters of galaxies are the most luminous extended sources in the extra-galactic X-ray sky, and therefore are easily detectable in large surveys. The importance of galaxy clusters in a cosmological context has been realized since the pioneering surveys undertaken with the \emph{Einstein} observatory \citep[e.g.][]{FormanJones1982,Gioia1990}, followed by studies with the ROSAT all-sky survey \citep[e.g.][]{ebeling2000, borgani2001, Boehringer2004, henry2009, bohringer2017, klein2017}. By simply counting the number of observed galaxy clusters one can confront cosmological model predictions and survey observations. However, it has been established that observational selection effects play a crucial role and must be controlled accurately when pursuing the goal of precision cosmology \citep[e.g.][]{vikhlinin2009, mantz2010, allen2011, Pacaud2016}.

X-ray astronomy will enter a new era with the extended ROentgen Survey with an Imaging Telescope Array \citep[eROSITA,][]{predehl2017}. This telescope is the primary instrument of the Russian/German Spektrum-Roentgen-Gamma (SRG) observatory, expected to be launched in 2019 (P.~Predehl, priv. comm.). eROSITA will possess unprecedented sensitivity and imaging capabilities for extended source emission \citep[][]{merloni2012}, and allow detecting $\sim 10^5$ galaxy clusters \citep[][]{Pillepich2012}. In order to detect this huge amount of galaxy clusters, eROSITA will scan the entire sky for four years, making it the second imaging X-ray all-sky survey ever made after ROSAT in the soft band ($0.5-2$~keV), and the first ever imaging survey in the hard band ($2-8$~keV). The promising capabilities of eROSITA bring great expectations to constrain dark matter and dark energy models through galaxy cluster science.

The derivation of a selection function for extended X-ray sources involves first their detection and then their classification as extended objects. Because extended objects are defined in contrast to point-like sources, this paper also focuses on the simulation and selection of point-like sources in the eROSITA All-Sky Survey (eRASS).

A reliable detection probability function of point-sources is crucial for assessing the completeness of samples, understanding the X-ray background, evaluating clustering studies, etc. Given the simple morphology of point-sources, detection probabilities may rely on knowledge of the local exposure time and background levels in a given observation \citep[e.g.][]{georgakakis2008}. An alternative and common approach consists in simulating mock observations accounting for a range of instrumental and astrophysical effects. Although this method is more computationally demanding, it embraces the entire chain from light emission to source detection and cataloguing, and this is the approach we adopted in this work.

As mentioned previously, a selection function for extended sources is a critical ingredient in almost all studies of the X-ray galaxy cluster population, including cosmological studies, scaling relation works \citep{stanek2006, Pacaud2007, mantz2010b, giodini2013,lovisari2015,andreon2016}, and detailed studies of the evolution of the intra-cluster medium physics and chemistry \citep[see][for a review]{BoehringerWerner2010}. The morphological complexity and diversity of the X-ray cluster population makes it more difficult to accurately describe selection effects. Comparison between samples detected at different wavelengths \citep[e.g.][]{wen2012,rozo2014,sadibekova2014,nurgaliev2017} allows an understanding of potential selection biases, but does not a priori provide a truth table for source detection. Therefore, Monte-Carlo simulations play an essential role in understanding the entire process leading to a validated galaxy cluster catalog. Reducing the diversity of cluster shapes to a sensible and reduced set of parameters sets limits on the computational demand and importantly, allows for a link between theoretical (e.g.~mass, redshift, etc.) and observational quantities. Cluster fluxes and apparent sizes are among the most relevant of these observables \citep{bohringer2000,Pacaud2006,burenin2007}.

Such synthetic simulations are not the unique route to address the selection function of clusters and AGN. 
Numerical N-body and hydrodynamic simulations play an increasingly important role in this debate. Indeed, as they become more and more realistic in reproducing the observed sky at multiple wavelengths \citep[e.g.][]{dolag2016}, they offer invaluable support in the understanding of selection biases. However, the still large computational requirements limit their usage for statistical studies.

The aim of this work is to forecast and illustrate realistic selection functions for the eRASS cluster and point-source population. It relies on multiple realizations of selected areas of the eROSITA sky, with X-ray emitting sources described by controlled parametric inputs. For instance, the galaxy cluster population is uniquely described by its apparent flux and size on the sky. We bring special attention into reproducing the main spectro-photometric features of the extragalactic point-source population (AGN). For the first time, we process eRASS simulation fields with the eROSITA source detection software (preliminary version). We derive realistic detection lists, similar to the real detection lists expected for scientific use. In particular, we explore thresholds needed to distinguish between spurious, point-like and extended sources and provide, given a chosen set of cuts, a first series of selection functions for point-like and extended sources. We demonstrate their practical usability with a prediction of the distribution of galaxy clusters in the eROSITA sky, by means of a forward-modelling approach.

This paper is constructed as follows. We first describe the built-in components in the simulations (Sect.~\ref{sect:simul_compo}), then we describe the simulation engine at the core of the analysis (Sect.~\ref{sect:simul_engine}).
We describe our selected simulation and instrumental setup, as well as our choice of fields in Sect.~\ref{sect:simul_setup}.
In Sect.~\ref{sect:simul_results} we show the source detection results, in particular the selection functions.
We discuss the impact of our important assumptions in Sect.~\ref{sect:simul_discu} and bring perspectives in Sect.~\ref{sect:simul_conclu}. 

Throughout the paper and unless stated otherwise, we assume a $\Lambda$CDM cosmology with $\Omega_\mathrm{m}=0.3$, $\Omega_{\Lambda}=0.7$ and $H_{0}=70$~km~s$^{-1}$~Mpc$^{-1}$.


\section{\label{sect:simul_compo}Simulated components}

This section presents the main expected components in a typical \emph{blank} field of our simulations of the extragalactic eROSITA sky.

	\subsection{AGN and cosmic X-ray background}
		\label{sect:sampling_lx}

We attempt to accurately reproduce the observed distribution of spectro-photometric properties of X-ray emitting active galactic nuclei (AGN). A list of spectra and positions, each corresponding to an individual source, is produced down to extremely low fluxes. The integration of the low-flux tail of the distribution provides a model for the unresolved X-ray background component up to the limit at which we simulate sources individually.

		\subsubsection{Spectral models}
		
We rely on a custom implementation of the formalism by \citet{gilli07} to generate spectral models on a log-spaced grid of energies in the range $[0.1,100]$~keV using \texttt{XSPEC} v12.7.0u \citep{arnaudxspec}. Parameters governing the spectral shape of a source are: a power-law photon index, $\Gamma$, the absorbing column density, $N_\mathrm{H}$, the source redshift, $z$, and the (unabsorbed) luminosity, $L_\mathrm{X}$, of the object in a given rest-frame 2 -- 10~keV band.
A critical parameter governing the choice of spectral model is the intrinsic absorption $N_\mathrm{H}$. We call {\it unobscured} those sources with $\log_{10} N_\mathrm{H}<21$, {\it Compton-thin} those showing $21<\log_{10} N_\mathrm{H}<24$, {\it Compton-thick mild} those with $24<\log_{10} N_\mathrm{H}<25$, and {\it Compton-thick heavy} those that have $\log_{10} N_\mathrm{H}>25$. For a given obscuration class, two regimes are considered, {\it Seyfert} or {\it QSO}, depending on whether the 0.5 -- 2~keV rest-frame luminosity of the source is lower or greater than $L_\mathrm{X}=10^{46}$~erg~s$^{-1}$.
We refer to \citet{gilli07} for details on the modeling of spectral energy distribution (SED) for each of these classes. The energy range and level of detail in the SED were chosen to match the expected detector performances of eROSITA. Depending on source class, they include a (cut-off) power-law with index $\Gamma$, and a $6.4$~keV iron line with various equivalent widths \citep{gilli99}, possibly modulated by a reflection component. \emph{Compton-thick mild} sources have their cut-off power-law replaced by a more complex \texttt{plcabs} model \citep{Yaqoob1997}. The source is redshifted before applying an additional absorption by the Galaxy ($N_\mathrm{H}^\mathrm{gal}$) depending on the location of the source on the sky. Finally, the flux of a source is obtained by integration of its SED, accounting for the luminosity distance computed in our reference cosmology.

		\subsubsection{Sampling the luminosity functions}
		
\begin{figure}
   \centering
   \includegraphics[width=\hsize]{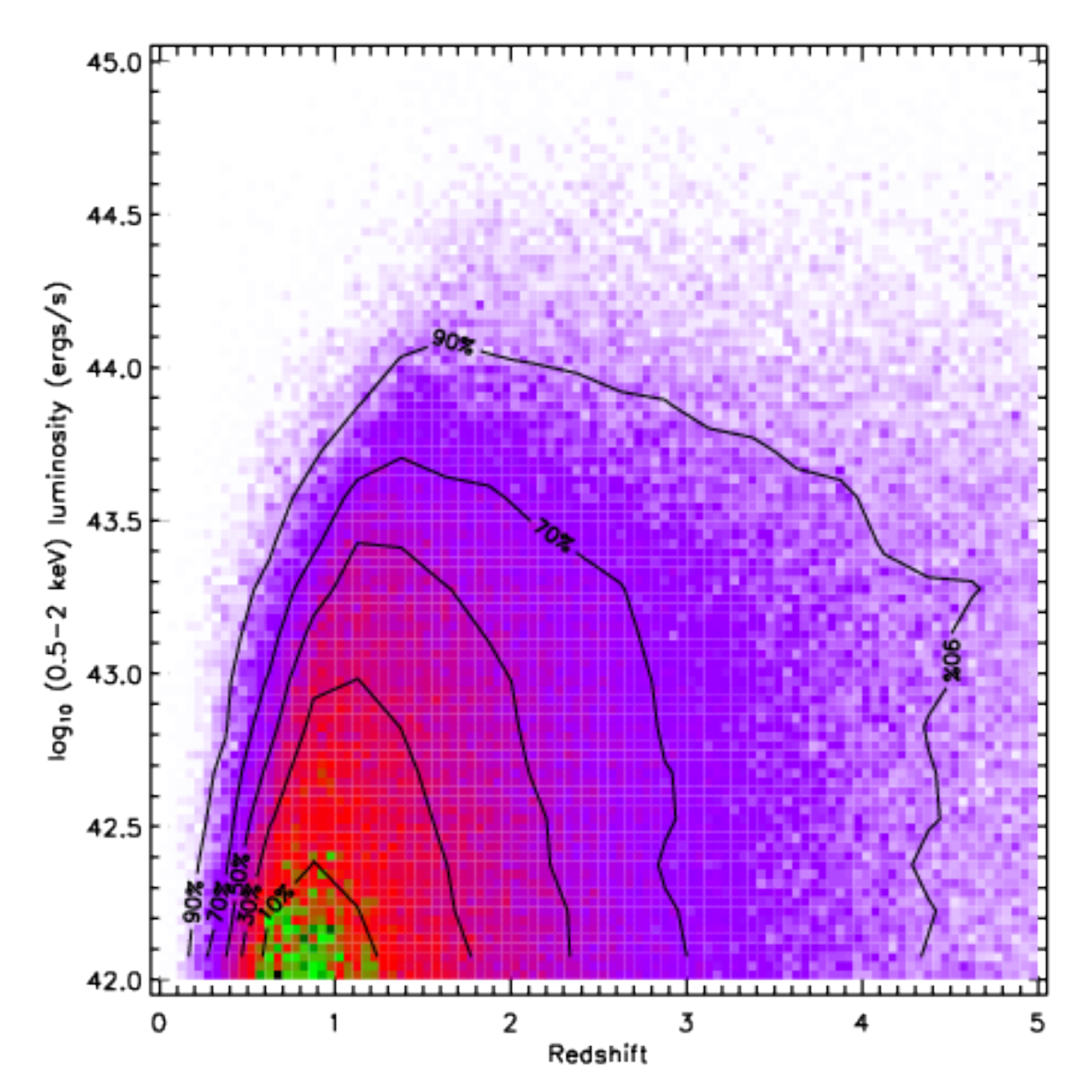}
      \caption{Two-dimensional histogram distribution of simulated sources in one realization of our X-ray AGN luminosity function sampling for a $22.7 \deg^2$ area on the sky ($253,297$ sources in total). Each black contour encloses the fraction of sources indicated as a label. To each source belongs one X-ray spectral model uniquely defined by the source luminosity, redshift, power-law index $\Gamma$ and absorbing column density $N_\mathrm{H}$ (Sect.~\ref{sect:sampling_lx}).}
         \label{fig:lxz_sample}
\end{figure}

Similarly to \citet{gilli07} we describe the luminosity function of unobscured AGN sources with the luminosity-dependent density evolution (LDDE) model of \citet{Hasinger2005}. Obscured sources are sampled from the LDDE modulated by a multiplicative factor, ranging from 4 to 1 as the source intrinsic luminosity increases. Obscuration values are distributed following the prescription by \citet{gilli07}, while power-law index parameters are drawn from a normal distribution of mean $\langle\Gamma\rangle=1.9$ and spread 0.2 regardless of the source obscuration level. Source luminosities range from $10^{42}$ erg s$^{-1}$ and redshifts span the $0<z<5$ interval. After accounting for the cosmological volume we compute the sky density $n(\Gamma,N_\mathrm{H},z,L_\mathrm{X})$ (units $\deg^{-2}$) and random-sample this distribution in order to obtain a discrete list of sources. Figure~\ref{fig:lxz_sample} represents the density of one such source list in the luminosity-redshift plane. Each source is then assigned a spectral energy distribution as described in the previous section. Sky positions are uniformly distributed in a field, as we do not aim to accurately model the spatial distribution of sources in this work (see Paper II, Ramos-Ceja et al. for a more detailed treatment).

\begin{figure}
   \centering
   \includegraphics[width=\hsize]{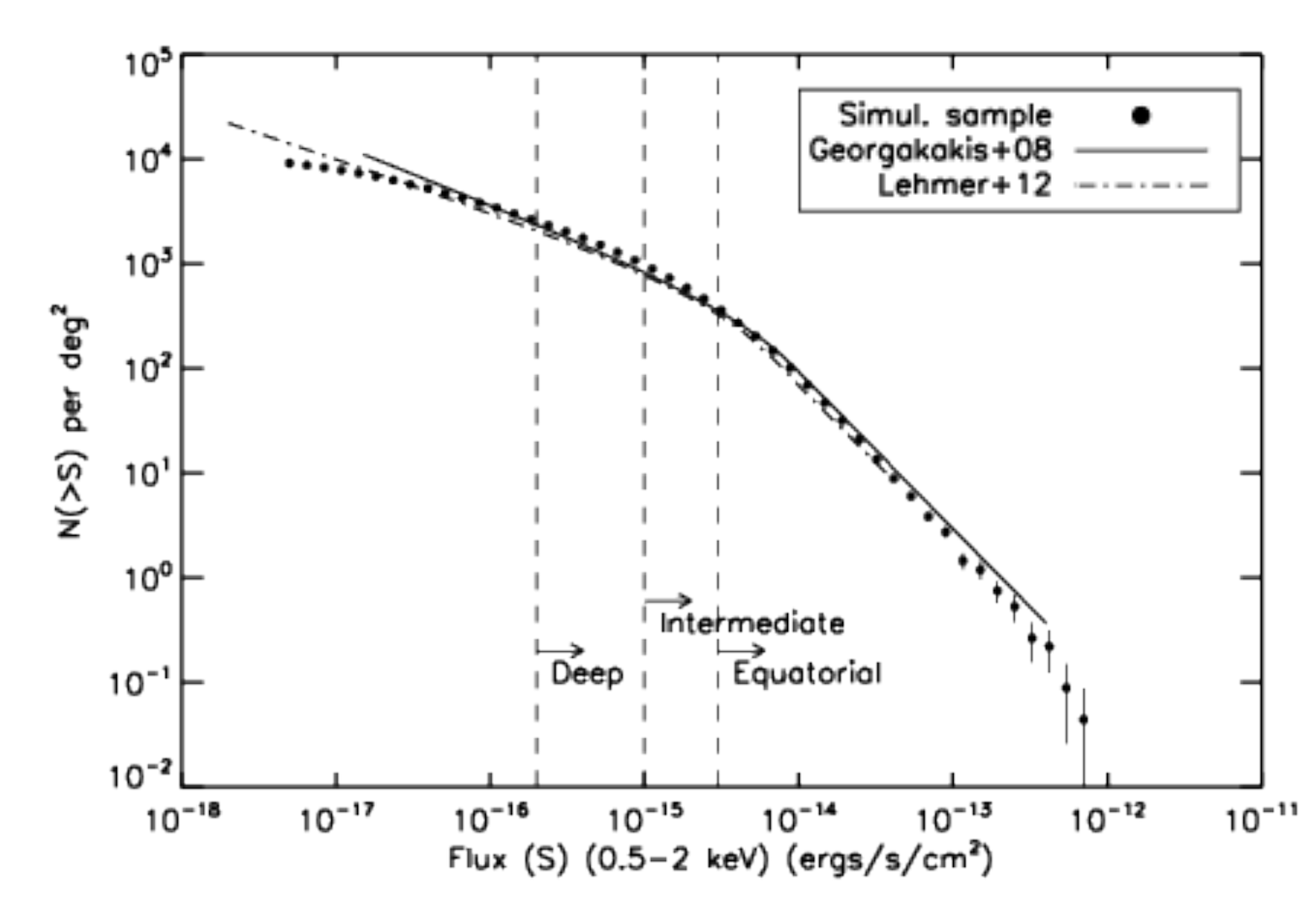}
      \caption{Soft-band cumulative source number counts for one realization of the X-ray AGN luminosity function sampling for a $22.7 \deg^2$ area on the sky. Error bars are $\sqrt n$ for each point. The parametrized $\log N-\log S$ from \citet{georgakakis2008} and \citet{lehmer2012} are overplotted (lines) for comparison. Vertical dashed lines indicate the flux $f_{\rm lim}$ of the faintest source being simulated in each of the three fields (Sect.~\ref{sect:sim_fields}).}
         \label{fig:lognlogs_sample}
\end{figure}

We verified the validity of our sampling procedure by computing the flux distributions of the simulated sources in different bands. We compared our results to \citet{gilli07} and to published $\log N-\log S$: the agreement in the soft-band is excellent (see Fig.~\ref{fig:lognlogs_sample}), while we predict twice as many heavily obscured sources ($\log_{10} N_H > 24$) in the 2 -- 10~keV band in comparison to \citet{gilli07}. We attribute this discrepancy for the rarest sources to our choices made in the high-energy modeling of the SED. This has practically no impact on this work which focuses on the soft-band characteristics of the eROSITA images.

		\subsubsection{Constructing the unresolved X-ray background}
		
\begin{figure}
   \centering
   \includegraphics[width=\hsize]{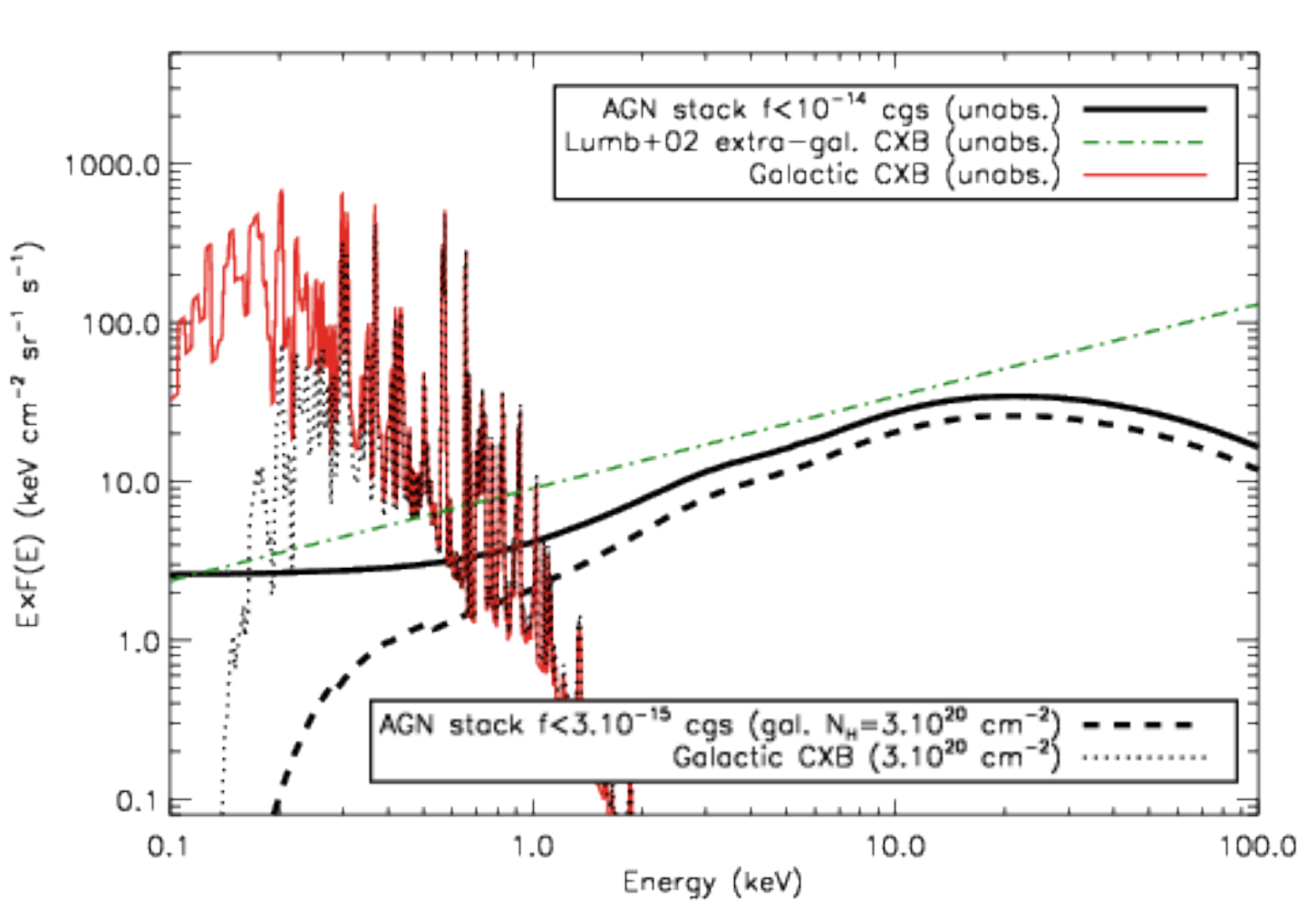}
      \caption{The energy spectrum of the simulated cosmic X-ray background components. The thick black dashed and plain lines are obtained with our model for AGN sources by stacking individual spectra of faint sources below $f_{\rm lim} = 10^{-14}$ (plain) or $3\times10^{-15}$ (dashed) erg\,s$^{-1}$\,cm$^{-2}$. For comparison, the dot-dashed green line shows the model of \citet{lumb2002} derived from {\it XMM-Newton} observations. Our emission model for the Galaxy (red and dotted lines) is described in the text.}
         \label{fig:all_cxb_backgrounds}
\end{figure}

The above-described procedure does not assume a lower limit on the flux of simulated sources. Sources well below the eROSITA detection limit are actually not simulated in order to save computation resources. A flux threshold $f_{\rm lim}$ is set depending on the exposure time of a simulated field (Sect.~\ref{sect:sim_fields}) and only sources with $f>f_{\rm lim}$ are individually simulated. The spectra of the remaining faint sources are stacked together and uniformly redistributed over a simulated patch of sky, hence constituting one single ``uniformly extended source'' instead of many point-sources. By doing so, we ensure self-consistent and realistic modeling of the spectral emission of the X-ray background (XRB) generated by unresolved AGNs. As an illustration, the spectrum of the AGN background component in the \textit{equatorial} field ($f_{\rm lim}=3 \times 10^{-15}$\,erg~s$^{-1}$~cm$^2$) in the 0.5 -- 2~keV band with galactic absorption $N_\mathrm{H}^\mathrm{gal} =3 \times 10^{20}$~cm$^{-2}$ is shown with a dashed line in Fig.~\ref{fig:all_cxb_backgrounds}. This figure also demonstrates the good agreement between the {\it XMM-Newton} measurements of \citet{lumb2002} (derived from {\it XMM-Newton} observations with sources excised down to $\sim 10^{-14}$ erg\,s$^{-1}$\,cm$^{-2}$ in the soft-band) and our unresolved XRB model with a similar $f_{\rm lim}$.

	\subsection{Extended sources as $\beta$-models}
	
Galaxy clusters are simulated in the simplest way using spherically-symmetric $\beta$-models \citep[][]{cavaliere1978} with different fluxes and core radii values and $\beta=2/3$. Our goal indeed is to derive selection functions that depend on a limited number of parameters. Sources representing galaxy clusters are randomly distributed across a simulated field, with a density of around $2$~per~deg$^2$.
Their spectral emission is rendered by an isothermal APEC model with $0.3~Z_{\odot}$ abundance, at temperature $T \in \{1, 5\}$~keV and redshift $z \in \{0.3, 0.8\}$. Clusters have 0.5 -- 2~keV fluxes chosen among discrete values ranging between $2\times 10^{-15}$ and $5 \times 10^{-13}$~erg~s$^{-1}$~cm$^{-2}$; core radii are also picked among discrete values ranging between $10$ and $80$~arcsec. The redshift and temperature of the spectral models have practically no impact on the 0.5 -- 2~keV energy conversion factors transforming fluxes into count-rates, hence no impact on the 0.5 -- 2~keV detection tests, which are the core of this study.

	\subsection{Particle and galactic background components}

In addition to the X-ray background originating from unresolved AGN in the field, two other main background components were added to our set of simulations. The contribution of unresolved galaxy clusters and groups to the eROSITA soft-band background is neglected, since it is a small component in the energy and sensitivity regimes relevant to this study \citep[e.g.][]{gilli99, gilli07, kolodzig2017}.

Following \citet{lumb2002}, the emission of the Galaxy is modeled with a double MEKAL model of temperatures $0.21$ and $0.081$~keV and solar abundance, representing the emission of the hot plasma located in the Galactic disk and halo. We assume a local photo absorbing column density equivalent to that of the field into consideration. We neglect here any spatially-dependent contribution to the Galactic background such as emission from the Hot Local Bubble.

Particle background is sampled from a list of events drawn from a GEANT4 simulation designed to reproduce the expected radiation environment at the Lagrange point L2 \citep{tenzer2010}. We assume this background component is not focused by the telescope mirror systems, hence it is not vignetted and impacts the detectors uniformly. Soft proton flares can create rapid enhancement in the level of unvignetted background. However, we limit our present study to the case of nominal particle background level and defer the analysis of the flare-induced background to further work. Therefore, the exposure assumptions in this work are on the optimistic side.

\begin{figure}
	\includegraphics[width=\hsize]{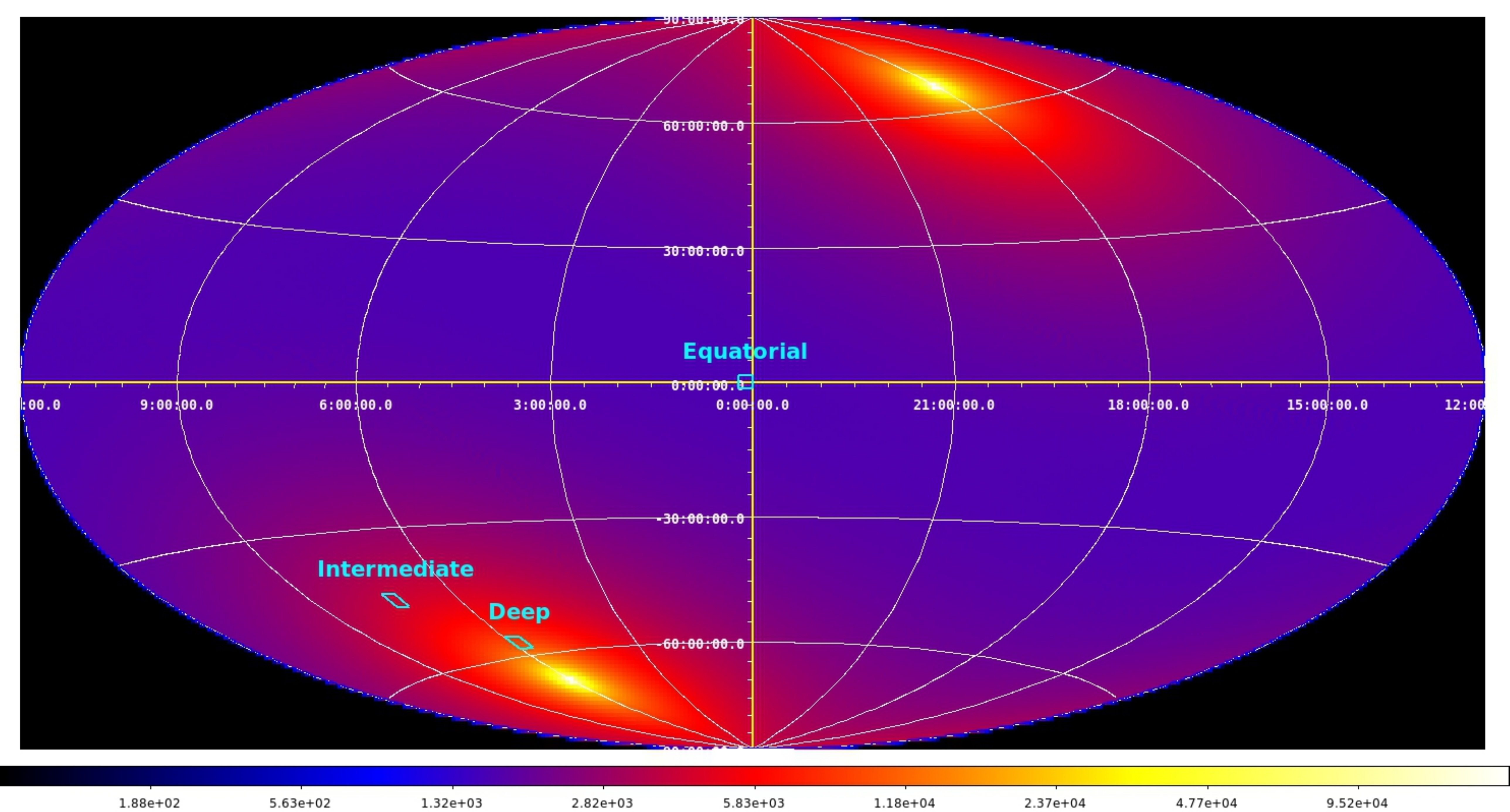}
      \caption{Simulated eROSITA all-sky 4-year exposure map in equatorial coordinates used in this work, with the location of the three relevant simulated fields: \emph{equatorial} ($\sim 2$~ks exposure time, uniform), \emph{intermediate} ($\sim 4$~ks, slight gradient), \emph{deep} ($\sim 10$~ks, larger gradient). The color bar (logarithmic scale) is in units of seconds.}
         \label{fig:syntheticfields_expo}
\end{figure}


\section{\label{sect:simul_engine}The eROSITA simulation engine}

The simulations presented in this paper result in realistic eROSITA calibrated event lists, similar to those expected to be delivered by the eROSITA ground-segment. Such event lists contain the arrival time, CCD coordinates, of the incoming events (photons or particles), as well as a reconstruction of their sky location and absolute energy. We reconstruct these characteristics assuming perfect knowledge of the calibration and spacecraft attitude.
We make use of the Monte-Carlo simulator SIXTE\footnote{http://www.sternwarte.uni-erlangen.de/research/sixte/}. It virtually implements a realistic transfer function converting sky photons into detector events, accurately accounting for CCD characteristics (including response functions and clocking) and telescope mirror behaviour. In order to save computation time, some parts of the telescope+instrument transfer function are modeled statistically, hence deviating from a pure ray-tracing simulator. These simplifications show notably at the mirror (point-spread and vignetting functions) and the CCD (response function) stages. We refer to \citet{schmid2012} for a detailed description of the SIXTE and its implementation in the context of eROSITA.

The detectors were simulated assuming an integration time of 50~ms and a finite readout time of the 384 CCD lines (pile-up effects are not relevant in this work). Response matrices are taken from rescaled EPIC-pn response matrices; those are of sufficient accuracy here, as we are focusing on broad-band properties.
The field-of-view of each of the 7 detectors is circular with a diameter $1.02$~deg, corresponding to the extent of the $384\times384$ pixel cameras with pixel size $9.6\arcsec$.	
	

\section{\label{sect:simul_setup}Instrumental and observational setup}

	\subsection{Exposure maps and attitude files}

A simple scanning strategy for the 4-year survey is assumed in this work, with the spacecraft scanning axis always pointing towards the Sun. The actual spacecraft law will be subject to subtle changes in the scanning pattern in order to fulfill angular constraints linked to, e.g.~the solar panels or stray-light requirements. Those ultimately lead to less uniform all-sky exposure maps, as discussed in \citet{merloni2012}. Since the present paper focuses on small patches of sky sufficiently far away from the ecliptic poles, these differences are neglected. Extrapolation of our results to the all-sky survey needs in principle a proper treatment of these exposure variations.
The corresponding attitude files describing the coordinates of the scanning axis in steps of $60$~s serve as input to the simulator. We assumed no gap nor jumps over the full duration of the survey, as well as ideal reconstruction of the attitude from the on-board star trackers.

	\subsection{Point-spread function and vignetting}
	
Photons originating from a source at infinite distance are redistributed using synthetic point-spread functions simulated with a ray-tracing procedure (P.~Friedrich, priv. comm.). It accurately reproduces an eROSITA ideal mirror system made of $54$~nested shells (Wolter-I configuration), including the spokes and the presence of an X-ray baffle. Such simulations were performed assuming a $1.6$~m focal length and a $0.4$~mm intra-focal shift of the detector relatively to the best on-axis focal point. This small shift was found to optimize the overall survey PSF size, at the cost of degrading the on-axis PSF. We note that the actual point-spread function will be measured on the sky when the instrument operates and compared to ray-tracing simulations and ground measurements (e.g. as done at the PANTER facility).

\begin{figure}
	\includegraphics[width=\hsize]{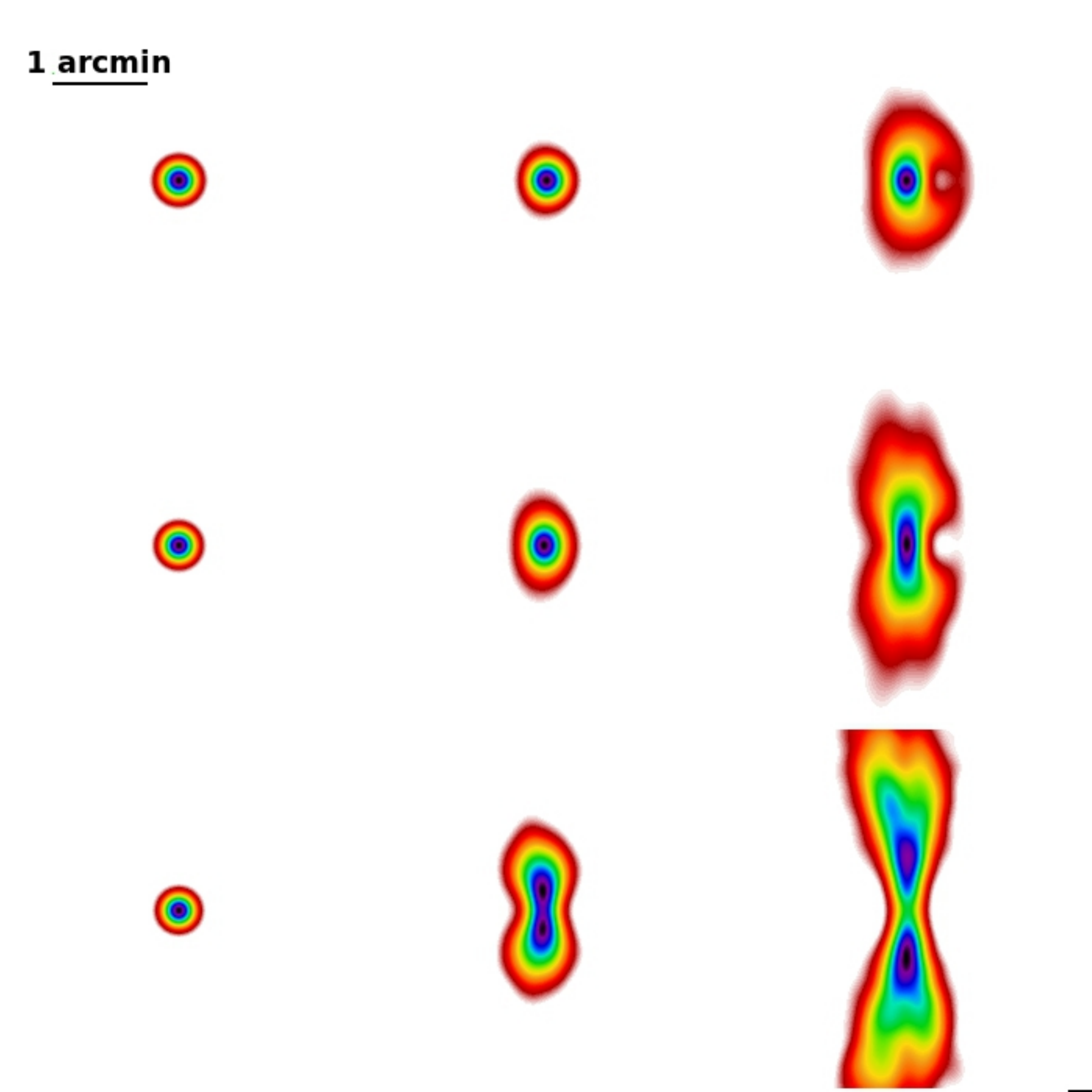}
      \caption{The ray-tracing simulated telescope point-spread function used in this paper. The images show the response of one eROSITA mirror module to a point-source at different incoming photon energies (from top to bottom: $1$, $3$ and $7$~keV) and different angular distances from the optical axis (from left to right: on-axis, 15$\arcmin$, 25$\arcmin$). The color scale in each panel is linear and encompasses the tenfold increase between the minimal (light red) and maximal (black) intensity, thereby emphasizing the typical shape distortions due to Wolter optics.}
         \label{fig:psf_grid}
\end{figure}

\begin{figure}
	\includegraphics[width=\hsize]{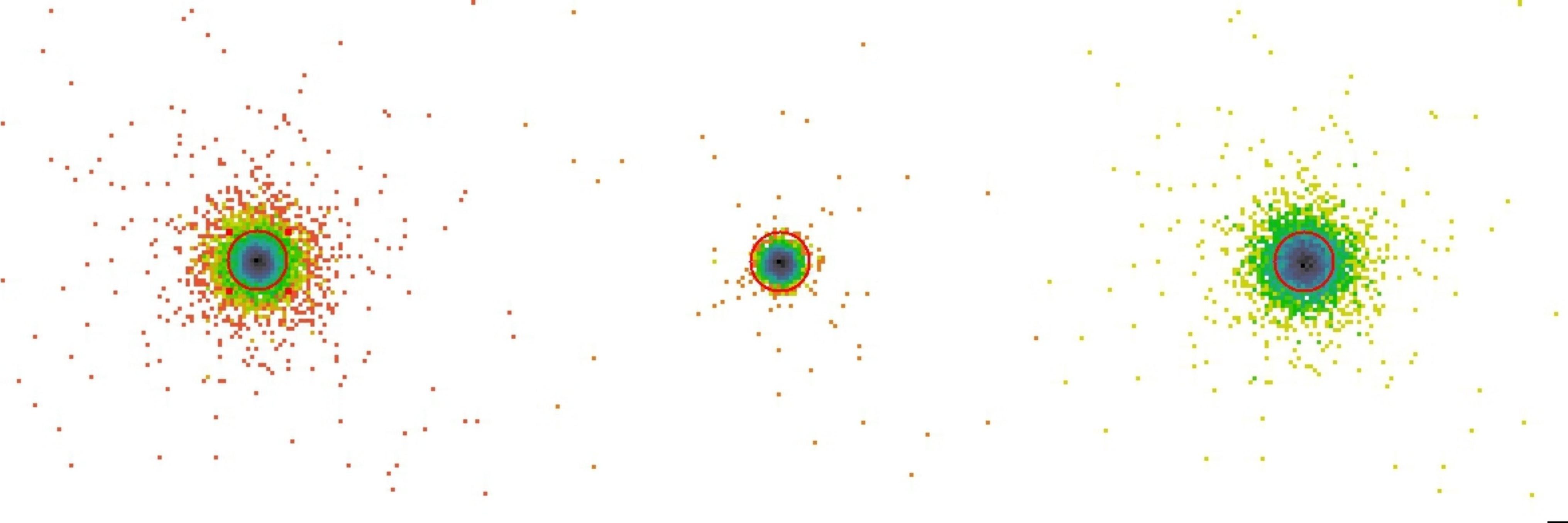}
	\includegraphics[width=\hsize]{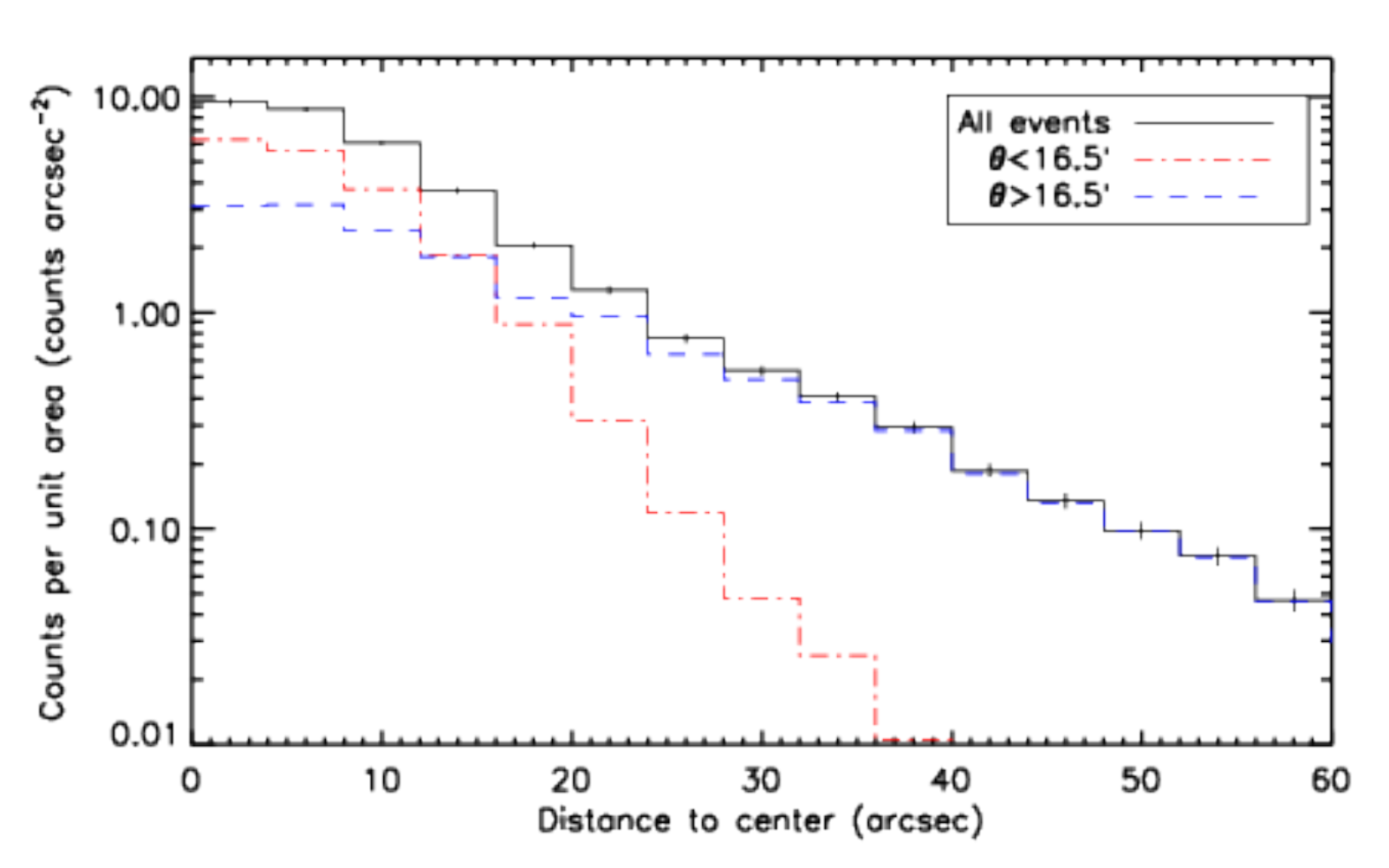}
      \caption{\textit{Top panel}: Simulation of a bright point-source with flux $10^{-11}$~erg~s$^{-1}$~cm$^{-2}$ in a 4-yr eROSITA equatorial region ($\sim 2$~ks exposure time). The image shows the sky projection of the 0.5 -- 2~keV source events collected by the seven CCD, binned with $4\arcsec$ pixels. Left: the ``survey PSF'', including all events. Middle: selecting only low off-axis events ($\theta < 16.5 \arcmin$, $40$\% of the total number of events). Right: selecting only large off-axis events ($\theta > 16.5 \arcmin$, $60$\%). The circle has a radius $30\arcsec$, slightly larger than the half-energy width of the survey PSF.
      \textit{Bottom panel}: The corresponding radial profiles in $4\arcsec$ bins (error bars are only shown for the top curve).}
         \label{fig:psf_all_in_out}
\end{figure}

The PSF we used is described as a tabulated series of images in steps of $1\arcmin$ off-axis angles ranging from $0$ to $30\arcmin$ and for energies $E=\{1,2,3,4,7\}$~keV (see Fig.~\ref{fig:psf_grid}). We assume constant PSF shape as a function of azimuthal angle, as we consider only axial rotation as is usual with the Wolter-I telescope symmetry.
Because it is counting photons individually, the ray-tracing simulation additionally provides an estimate of the vignetting factor on a grid of energies and off-axis angles. It is used to compute the ratio of flux between double-reflected photons and all photons emitted by a source located at a given off-axis angle, and usually expressed relative to the on-axis position. Fig.~\ref{fig:psf_all_in_out} shows the combined effect of vignetting and PSF distortion on a bright point source passing about $\sim 50$ times through the eROSITA field-of-view during the 4-year scan duration.
		
\begin{figure*}
	\includegraphics[width=\linewidth]{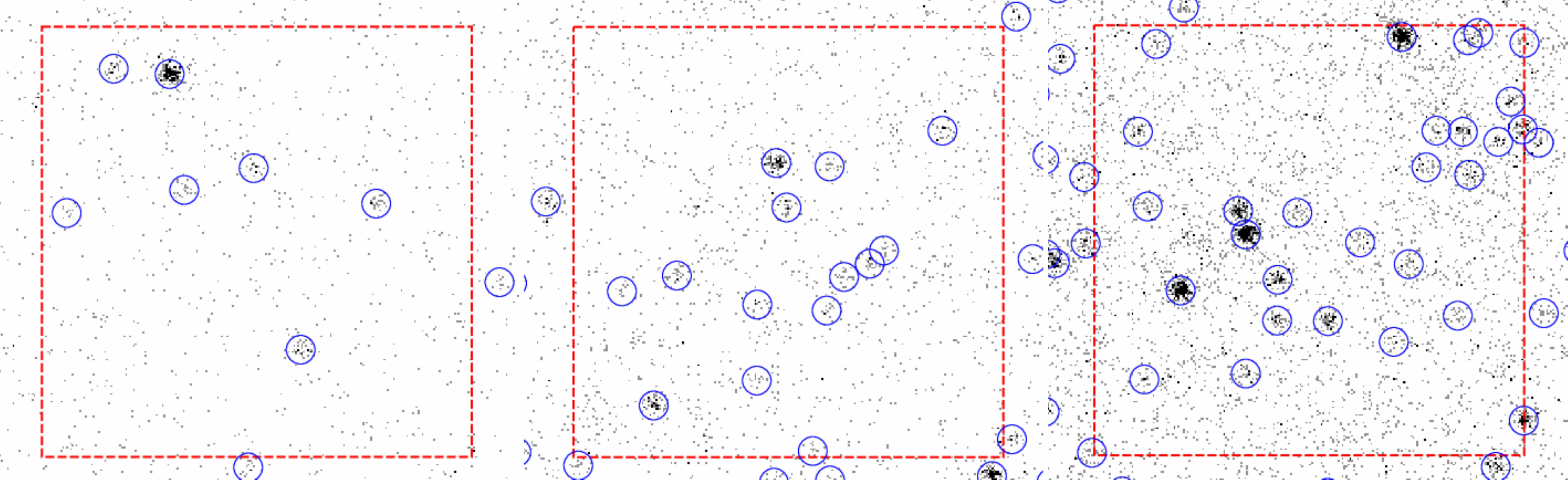}
      \caption{Zoom over three simulated eROSITA extragalactic survey fields (\emph{equatorial}, \emph{intermediate} and \emph{deep}, from left to right) in the 0.5 -- 2~keV band, free from galaxy clusters (i.e.~containing only backgrounds and AGN as point-sources). North is up and East left, each dashed square is $15\arcmin$ on a side. The blue circles have a radius of $0.5\arcmin$ and show the position of the detected sources. The pixel scale is $4~\arcsec$ and identical grey scales are applied to each image to emphasize the differences in sensitivity.}
         \label{fig:syntheticfields_images_zoom}
\end{figure*}
		
	\subsection{Simulated fields}	
	\label{sect:sim_fields}
	
We selected three fields at specific locations in the eROSITA sky (see Fig.~\ref{fig:syntheticfields_expo}). A field corresponds to an elementary region of the eROSITA sky tiling pattern, and shows as a $3.6 \deg \times 3.6 \deg$ square in tangential projection. In the following we name these fields: \emph{equatorial} ($\sim 2$~ks exposure time, uniform), \emph{intermediate} ($\sim 4$~ks, less uniform) and \emph{deep} ($\sim 10$~ks, larger exposure gradient). Table~\ref{tab:field_description} provides key parameters relevant to these simulated fields. 
	
\begin{table}
 \centering
  \begin{tabular}{lccc}
  \hline
  \hline
  & Equatorial & Intermediate & Deep\\
\hline
$T_\textrm{exp}$    &   $\sim 1.6$ ks &   $\sim 4$ ks   &   $\sim 9.7$ ks  \\
$\Delta T_\textrm{exp}$ max.   & 0 \%  &   16 \%  &   33 \%  \\
$N_\mathrm{H}^\mathrm{gal}$ (cm$^{-2}$) &   $3 \times 10^{20}$  &   $8.8 \times 10^{20}$    &   $6.3 \times 10^{20}$ \\
$f_{\rm lim}$ (cgs)  &   $3 \times 10^{-15}$ &   $10^{-15}$  &   $2 \times 10^{-16}$ \\
\hline
  \end{tabular}
  \caption[Description of the simulated fields]{Global parameters for the three types of fields simulated in this study. Each field is a square of $3.6 \deg$ on a side. The galactic absorption is assumed uniform with a value $N_\mathrm{H}^\mathrm{gal}$. AGN are simulated individually down to a flux $f_{\rm lim}$ in the 0.5 -- 2~keV band and sources below $f_{\rm lim}$ contribute to a diffuse background component. The maximal variation of exposure across a field is listed as $\Delta T_\textrm{exp} = (T_\textrm{max}-T_\textrm{min})/T_\textrm{mean}$.}
 \label{tab:field_description}
\end{table}


	\subsection{Images of the synthetic fields}

Fig.~\ref{fig:syntheticfields_images_zoom} shows a $15\arcmin \times 15\arcmin$ excerpt of the images created out of the simulated event lists of \emph{blank fields}, i.e.~fields without galaxy clusters.
The distribution of point-like sources is uniform over the sky: any slight apparent gradient in source concentration is an effect of varying exposure times across the fields. The increase in sensitivity clearly makes more sources visible by eye; this figure also outlines the excellent angular resolution of eROSITA, well-adapted to beat confusion effects over most of the survey area, even in deep fields.


\section{\label{sect:simul_results}Results}

	\subsection{Source detection and characterization}

The source detection and characterization procedure used in this work is a preliminary version of the source detection tool in the eROSITA Science Analysis Software System (eSASS) package. It builds upon the source detection algorithm used in the \textit{XMM-Newton} Science Analysis System (XMM-SAS) with several revisions and upgrades. The detection procedure is based on the sliding-cell method. As a first step, this algorithm scans an X-ray image with a sliding square box, and if the signal-to-noise in the box is greater than a specified threshold value it is marked as a source candidate. The signal is calculated from the pixel values inside the cell, and the background is estimated from the neighboring pixels. Then, the candidate objects are removed from the image creating a source-free image which is interpolated by a spline function to create a smooth background map. The algorithm convolves the input image with a $9\times9$~pixel ($36\times36$~arcsec) kernel described by a $\beta=2/3$-profile with $r_\mathrm{c}=15$~arcsec, which roughly matches the survey PSF. The convolved image and the corresponding background map are then used to calculate a signal-to-noise map, in which the significant peaks are the positions of the detected sources. In order to increase the sensitivity for large extended sources, this procedure is repeated for $2\times2$ and $4\times4$ rebinned images corresponding to kernels with $r_\mathrm{c}=30$ and $r_\mathrm{c}=60$~arcsec, respectively. 

Each source candidate identified by the sliding cell algorithm is further analyzed by a maximum likelihood fitting method. This technique compares the spatial distribution of the input sources with a PSF\footnote{This PSF is based on the ray-tracing simulations with $0.4$~mm focus offset (see Section~\ref{sect:simul_setup}).} convolved with a source extent model ($\beta$-profile). The final log-likelihood is calculated by varying the input source parameters, i.e. position, counts, extent. A multi-PSF fit is also implemented which helps in deblending and reconstructing the parameters of close by sources. In the output list, only sources with a log-likelihood above a given threshold are kept.

Among the maximum likelihood fit output parameters of interest are: \textit{i}) \textit{detection log-likelihood} gives the significance of the detection; \textit{ii}) \textit{extent} is the apparent extension of the best fitting $\beta$-model in pixel units; and \textit{iii}) \textit{extension log-likelihood} compares the significance of the extended model and the point-like model. This last parameter basically classifies the detected sources as point-like (value equal zero) or as extended-like (value greater than zero).

Given that the PSF fitting of the maximum likelihood fitting method is more sensitive to the core of the PSF when on- and off-axis photons are separated, two images, from the same simulation and covering the same sky region, are produced with photons chosen according to their position on the FoV. The photons are split into inner photons ($<16.5'$) and outer photons ($>16.5'$). In this way, the source detection pipeline runs simultaneously over two images (see Fig.~\ref{fig:psf_all_in_out}).

All simulated images were analyzed with the method described above. The detected sources were cross-identified with the simulation inputs using a matching radius of $28$~arcsec for point-like sources and $80$~arcsec for extended ones.

\begin{figure}
	\includegraphics[width=\hsize]{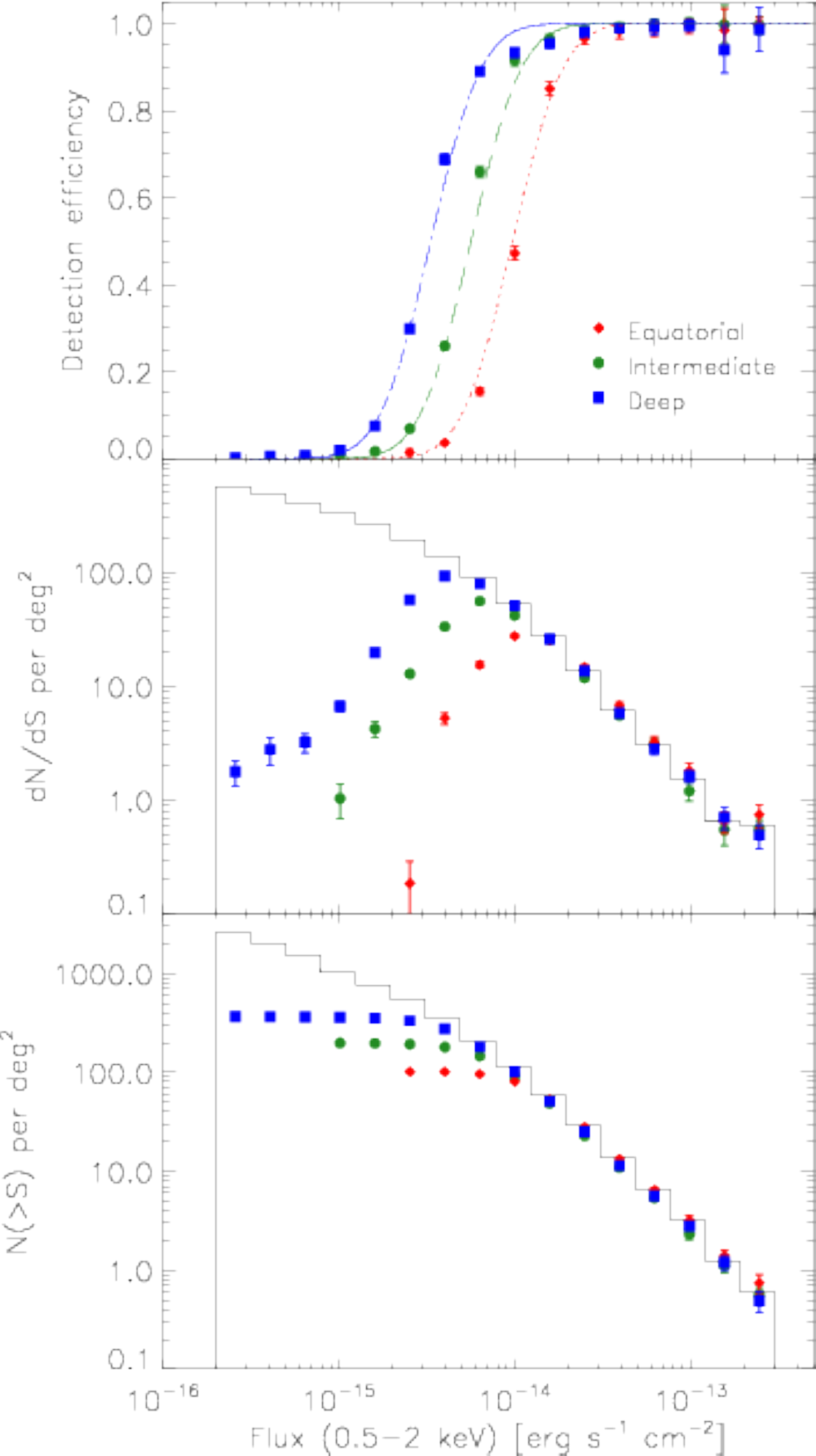}
      \caption{Point-like source completeness analysis for all three simulated sky regions: {\it Equatorial} (red diamonds), {\it Intermediate} (green circles) and {\it Deep} (blue squares). The abscissa is the input source flux. \textit{Top panel}: Point-like detection efficiency. Discontinuous lines represent the parametrized models described in App.~\ref{sect:modelfits}. \textit{Middle panel}: Differential number counts. \textit{Bottom panel}: Integral number of point-like sources. In the \textit{middle and bottom panels} the solid line shows the input distribution. The error is given by the standard deviation over the simulations.}
         \label{fig:agnsel}
\end{figure}

    \subsection{Source classification}
    \label{subsec:src_classification}

A trade-off between sample completeness and contamination is inevitable when the source selection function in surveys is estimated. Following a methodology introduced in \citet{Pacaud2006} we explore the output parameter space of the maximum likelihood fitting method by means of our simulations in order to set point-like and extended source classification criteria and to estimate their contamination by spurious and misclassified sources. We call spurious detections those that cannot be identified with any input source within the search radius, and misclassified sources to those point-sources classified by the pipeline as extended sources or vice-versa. We will refer to false detections as a single concept that includes spurious and misclassified detections.

        \subsubsection{Point-source selection functions}

AGNs represent the dominant extra-galactic population at X-ray wavelengths. Although the goal of this work is to determine the galaxy cluster selection function, the estimation of the point-like detection efficiency and its contamination helps to control the systematics in the detection and characterization of the extended source population.

First, we restrict ourselves to estimate the false detection rate based on the \emph{blank} field simulations, i.e.~with point-like sources plus background only. We simulate $30$~times each field. We found that a simple threshold in the source detection log-likelihood parameter removes most of the false point-like sources while maintaining a good detection efficiency. We choose a threshold value of $10$, obtaining $\sim 0.1$,~$\sim 0.2$, and $\sim 1.1$~spurious~sources~per~deg$^2$ for the \emph{equatorial}, \emph{intermediate} and \emph{deep} fields, respectively. Such false detection numbers correspond to $\sim 0.1\%$, $\sim 0.2\%$ and $\sim 0.3\%$ of the average detected sources~per~deg$^2$ in their respective field.

The resulting AGN detection efficiency as a function of input flux is shown in the top panel of Fig.~\ref{fig:agnsel}. This efficiency is obtained by calculating the ratio of the cross-identified objects to the input sources. The displayed error is given by the standard deviation over the $30$ simulations of each simulated field. For the \emph{equatorial} field, the point-like sources have a $90\%$ completeness at a flux limit of $\sim~1.7\times10^{-14}$~erg~s$^{-1}$~cm$^{-2}$, while for the \emph{intermediate} field this flux limit is $\sim9.7\times10^{-15}$~erg~s$^{-1}$~cm$^{-2}$, and for the \emph{deep} field it is $\sim6.5\times10^{-15}$~erg~s$^{-1}$~cm$^{-2}$. The $50\%$ completeness is reached at $\sim~1.0\times10^{-14}$~erg~s$^{-1}$~cm$^{-2}$ for the \emph{equatorial} field, while for the \emph{intermediate} field this flux limit is $\sim5.2\times10^{-15}$~erg~s$^{-1}$~cm$^{-2}$, and for the \emph{deep} field it is $\sim3.1\times10^{-15}$~erg~s$^{-1}$~cm$^{-2}$. The large error bars in bright sources reflect mainly their lower number density, which is given by the AGN $\log N-\log S$ distribution.

\begin{table*}
 \centering
  \begin{tabular}{ | l | c || c || c |}
  \cline{2-4}
   \multicolumn{1}{c|}{} & Equatorial field & Intermediate field & Deep field \\
   \multicolumn{1}{c|}{} & [sources per deg$^2$] & [sources per deg$^2$] & [sources per deg$^2$] \\
  \hline
   Spurious extended sources & $0.04\pm0.06$ & $0.05\pm0.09$ & $0.13\pm0.09$ \\
   Spurious point-like sources & $0.11\pm0.08$ & $0.19\pm0.10$ & $1.00\pm0.30$ \\
   Misclassified point-like sources & $0.44\pm0.09$ & $1.38\pm0.20$ & $8.40\pm1.80$ \\
  \hline
 \end{tabular}
  \caption[Number of false detections in point-like and extended simulations]{Number of spurious and misclassified extended sources (galaxy clusters) and point-like sources (AGN) in the \emph{cluster} field simulations on the \emph{equatorial}, \emph{intermediate} and \emph{deep} fields.}
 \label{tab:spurious_sources_eSASS}
\end{table*}

\begin{figure*}
	\includegraphics[width=\textwidth]{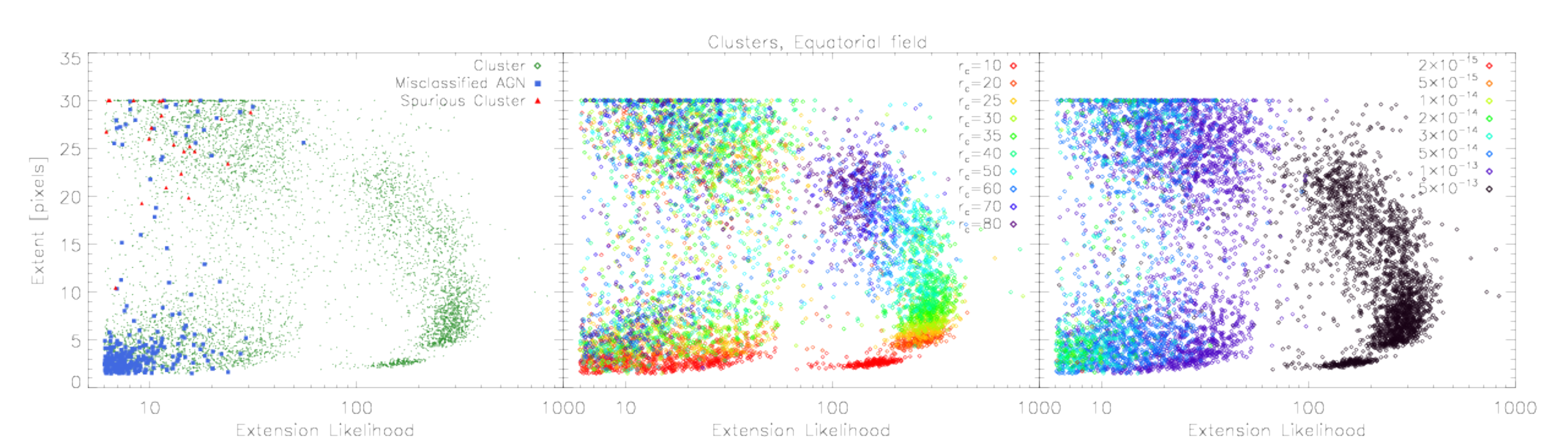}\\
	\includegraphics[width=\textwidth]{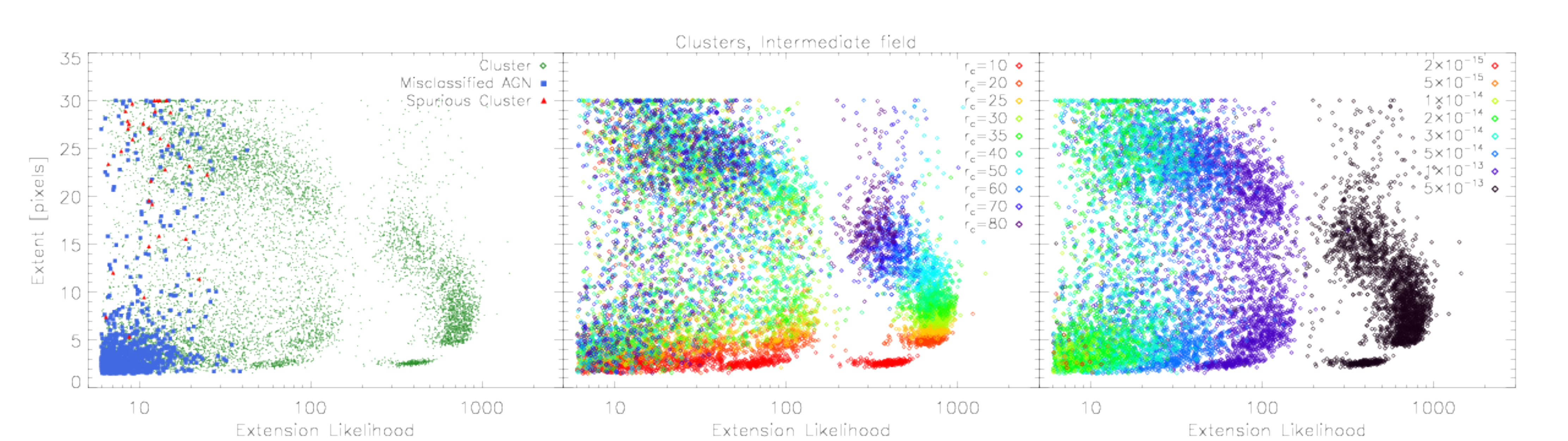}\\
	\includegraphics[width=\textwidth]{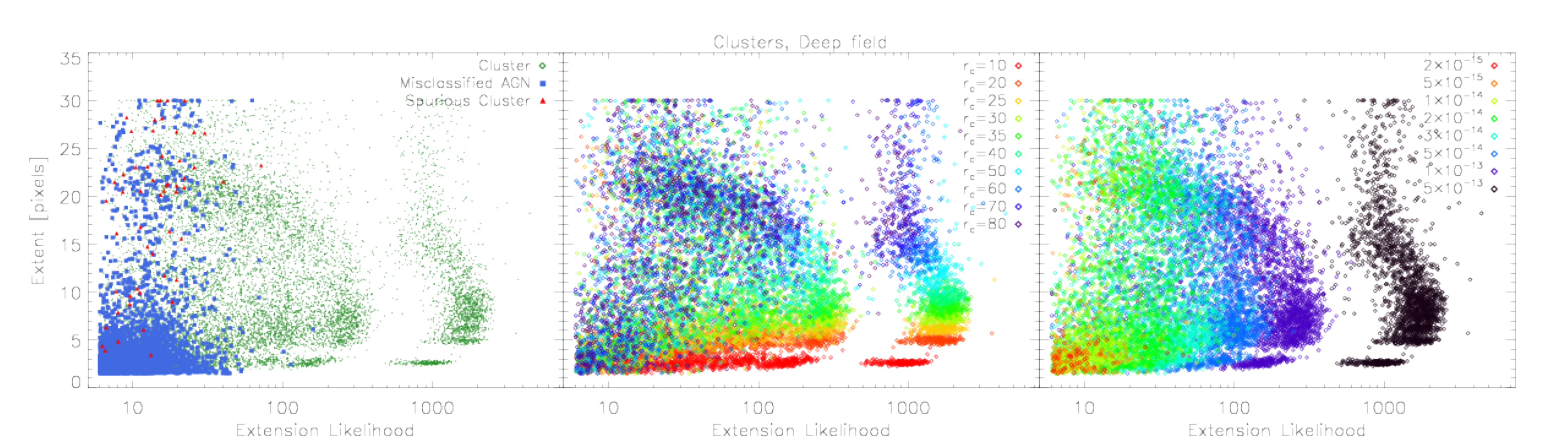}
      \caption{
      Final selection criteria for extended sources, from the preliminary version of the eSASS pipeline, with optimal (low-contamination) parameters. The extent - extension log-likelihood plane is shown for the three simulated sky fields: \emph{equatorial} (\textit{top}), \emph{intermediate} (\textit{middle}) and \emph{deep} (\textit{bottom}). \textit{Left panels}: simulated (and detected) clusters are displayed as green dots, spurious extended detections as red triangles, and AGN classified as extended sources in blue squares. \textit{Middle panels}: Only input detected galaxy clusters are displayed (green diamonds in the \textit{left panels}). The distinct colours show the different simulated core radii (in arcsec). \textit{Right panels}: Only input detected galaxy clusters are displayed. The different colours show the distinct simulated input fluxes (in units of erg~s$^{-1}$~cm$^{-2}$).}
         \label{fig:clu_ext_extlk}
\end{figure*}

	    \subsubsection{Cluster selection functions}

The extended source classification is a complicated task since it has not only to deal with spurious detections but also with misclassified point-like sources, i.e. point-like sources characterized as extended. Moreover, extended sources usually have a low surface brightness making their detection and characterization a difficult process. Our goal is to find a location in the detection/characterization parameter space where the majority of the simulated extended sources are recovered while keeping the contamination level at a reasonable rate. This is of special importance given that the goal of eROSITA is to use galaxy cluster counts to constrain the dark energy. We remind here that in contrast with the AGN population, which was simulated following a $\log N-\log S$, sources representing galaxy clusters are randomly distributed across the simulated fields with a density of around $2$~per~deg$^2$ (see Section~\ref{sect:simul_compo}).

Besides the source detection log-likelihood values stated in the previous section, we scanned the source extent - extension log-likelihood parameter space to look for criteria that allow us to obtain a large and uncontaminated extended source sample while maintaining a high detection rate. For this, we use \emph{cluster} fields, i.e.~simulations that contain X-ray background, point-like and extended sources. Fig.~\ref{fig:clu_ext_extlk} shows the final selection process in the extent - extension log-likelihood plane for the \emph{equatorial} (top), \emph{intermediate} (middle) and \emph{deep} (bottom) fields.

We specify that the maximum extent value that the algorithm should assign to a source is $30$~pixels ($120$~arcsec), even if the algorithm drifts towards a larger value. The minimum requested extent value is $1.5$~pixels ($6$~arcsec), and the threshold of the extension log-likelihood is $6$. These thresholds ensure a low contamination by spurious sources, but the number of misclassified point-like sources varies in the different fields. For the \emph{equatorial} field we obtain $\sim 0.5$~false~extended~sources~per~deg$^2$. In the \emph{intermediate} field we have $\sim 1.4$~false~extended~sources~per~deg$^2$, and for the \emph{deep} field we obtain $\sim 8.5$~false~extended~sources~per~deg$^2$.

Table~\ref{tab:spurious_sources_eSASS} shows in detail the fraction of spurious and misclassified sources in each simulated field. It is worth mentioning that similar numbers of spurious and misclassified sources are found in both the {\it blank} and {\it cluster} fields when using the same thresholds. In Section~\ref{subsec:cosmo_forecast} we forecast the number of expected clusters assuming a survey with a depth equal to the \emph{equatorial} field all over the sky. We expect to detect $\sim 5.2$~clusters~per~deg$^2$ plus $10\%$ of contamination, i.e. from our false sources.

The middle and right panels of Fig.~\ref{fig:clu_ext_extlk} show the extended sources color-coded according to the input core radius and flux values, respectively. The middle panels display the distribution of the discrete values used for the core radius of the simulated clusters (see Section~\ref{sect:simul_compo}), while the right panels show that mainly sources with high-flux end within the plane of the selection criteria.

As seen in Fig.~\ref{fig:clu_ext_extlk}, one could put more stringent criteria to obtain a non-contaminated cluster sample, e.g. increasing the minimum value of the extension log-likelihood, but this would lead to excluding a considerable amount of extended sources, especially the faintest ones.

\begin{figure}
	\includegraphics[width=\hsize,trim=0 0 0 1.2cm,clip]{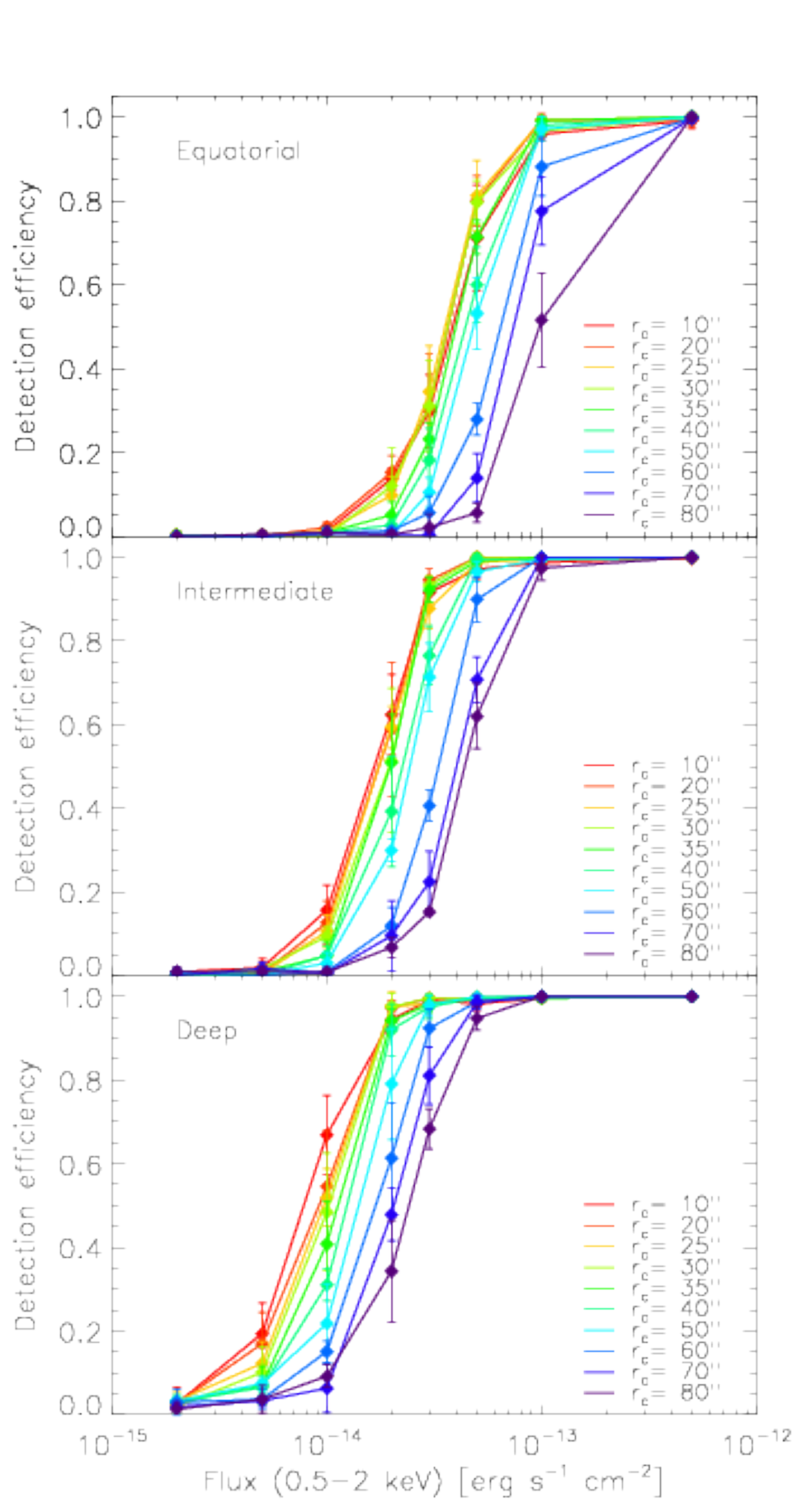}
      \caption{Extended source detection efficiency from the eSASS pipeline in the \emph{equatorial} ($\sim2$~ks exposure, \textit{top}), \emph{intermediate} ($\sim4$~ks, \textit{middle}) and \emph{deep} ($\sim 10$~ks, \textit{bottom}) simulated fields as a function of input flux and for each simulated core radius value.}
         \label{fig:cludeteff}
\end{figure}

\begin{figure}
	\includegraphics[width=\hsize]{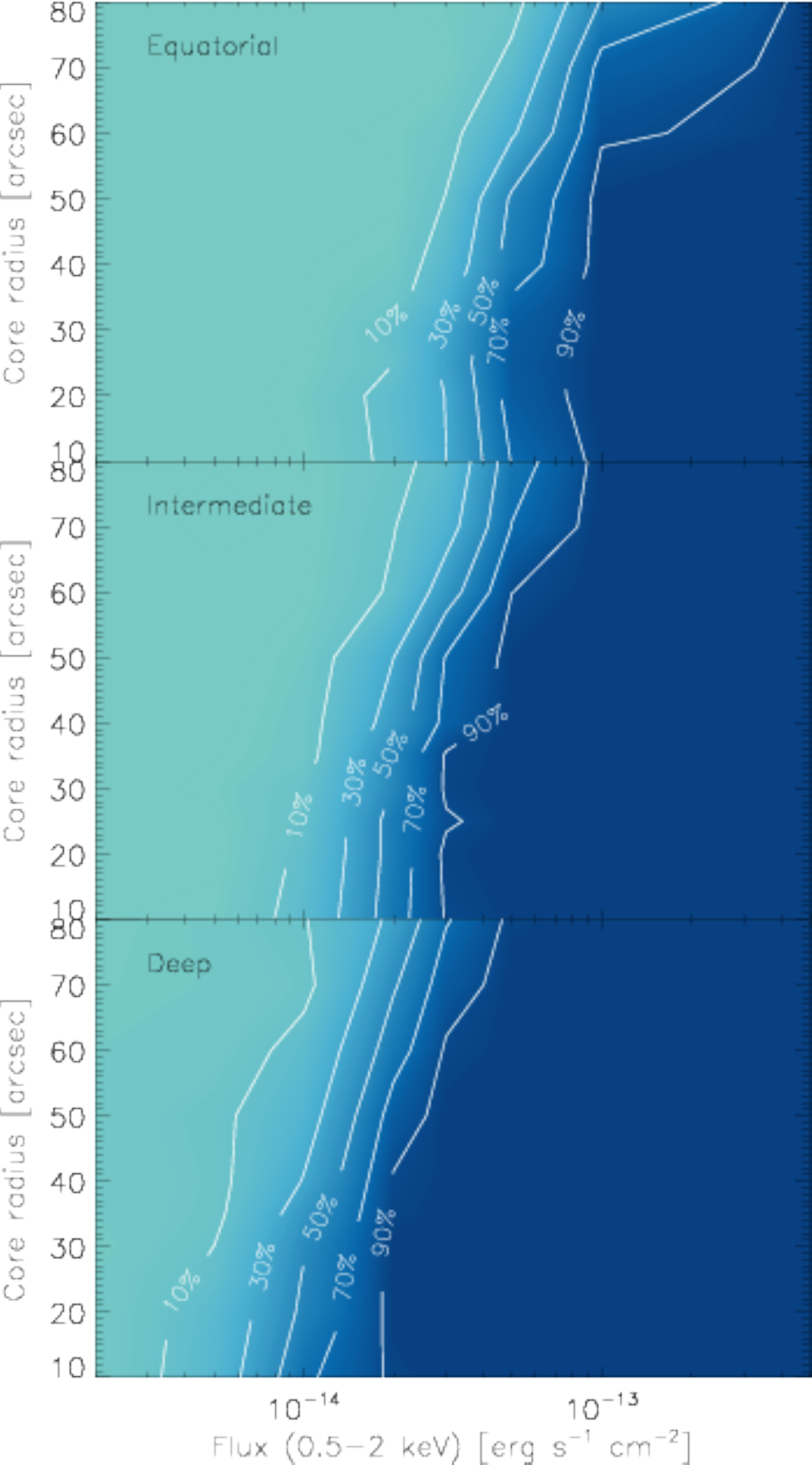}
      \caption{Extended source detection efficiency of the eSASS pipeline in the \emph{equatorial} ($\sim2$~ks exposure, \textit{top}), \emph{intermediate} ($\sim4$~ks, \textit{middle}) and \emph{Deep} ($\sim 10$~ks, \textit{bottom}) simulated fields as a function of input flux and core radius.}
         \label{fig:clusel}
\end{figure}

The normalized detection probabilities of extended sources for the three simulated fields are presented in Fig.~\ref{fig:cludeteff}, as a function of the input flux. In these plots, detection efficiency equals $1$ means that $100\%$ of the simulated sources have been detected and classified as extended. As expected, the deeper is the observation the fainter extended sources are recovered. Fig.~\ref{fig:clusel} shows the mean detection probability of extended sources as a function of input flux and input core radius. Similar to other works \citep[e.g.][]{vikhlinin1998,Pacaud2006,Clerc2012} we also found that the extended source detection efficiency is not a function of source flux only, especially for the shallower observations.

\section{\label{sect:simul_discu}Discussion}

	\subsection{Effect of source classification criteria}
	\label{subsec:impact_srcclass}
	
One could argue that the number of false extended source detections, i.e. spurious and misclassified detections, found in the different simulated fields (see Table~\ref{tab:spurious_sources_eSASS}) is high considering that eROSITA will perform an all-sky survey. However, most of the false extended detections are misclassified point-sources. Such sources might be close pairs of point-sources which cannot be disentangled by the detection algorithm and were therefore classified as an extended source. One way to reduce the number of misclassified sources is by doing a complete follow-up on the detected extended sources. Another way is by putting stricter thresholds in source classification criteria, e.g. by increasing the extent and extension log-likelihood thresholds (see Section~\ref{subsec:src_classification} and Fig.~\ref{fig:clu_ext_extlk}). For example, using a threshold value in extension log-likelihood of $20$ reduces by $\gtrsim 95\%$ the number of missclassified point-like sources in the three fields. Although such an approach gives a cleaner sample, many real extended sources are missed.

	\subsection{Relevance on cosmological forecasts}
	\label{subsec:cosmo_forecast}

Uncertainties in the selection function of a sample of clusters can introduce biases to the cosmological constraints which are determined from them. In this section we discuss the impact that incomplete knowledge of the selection has on the recovered cosmological constraints. For this test we follow the methodology of \citet{Clerc2012} and utilise the z-CR-HR method. We assume that the selection has eliminated all spurious clusters and misclassified AGN.

\subsubsection{The z-CR-HR method}
The z-CR-HR method is based on the premise that the raw X-ray data of a galaxy cluster contains significant information about its redshift, luminosity and temperature and that this information can be statistically extracted. The cosmological analysis is then simplified by basing it on purely the cluster redshift and quantities which are directly observable in X-rays, namely the count-rate in the 0.5 -- 2~keV band (CR) and the hardness ratio (HR), which is the ratio of the count-rates measured in the 1 --2~keV and 0.5 -- 1~keV bands. A particular advantage of this method is that it bypasses the need of having to derive individual cluster masses, X-ray luminosities and temperatures and that the scaling relations between mass and its X-ray proxies can be constrained simultaneously with the cosmological parameters. The key steps in this procedure are as follows:
\begin{itemize}
\item Compute the halo mass function;
\item Derive the 3D distributions of temperature, luminosity and core radius using the $M-T$, $L-T$ and $M-r_c$ scaling relations, taking the relevant scatters into account;
\item Apply an instrumental model for eROSITA to obtain a theoretical distribution of clusters in the CR-HR plane for each slice in the redshift space;
\item Apply the selection function to obtain a synthetic observed distribution of clusters that one would expect eROSITA to detect (here the {\it equatorial} selection for nominal thresholds, Fig.~\ref{fig:cludeteff}, top);
\item Apply an error model to account for measurement errors of CR and HR.
\end{itemize}

\subsubsection{Simulated eROSITA z-CR-HR catalogues}
After following the procedure described in the previous section and with the un-convolved, error-free z-CR-HR distribution in hand, we randomly sampled the CR-HR plane for each redshift slice to obtain a catalogue of mock clusters each with a redshift, count-rate and hardness ratio. A Gaussian random error of 10\% and 20\% for CR and HR respectively is then added to each cluster in the mock catalogue\footnote{Although the characterization of photometric measurements is beyond the scope of this paper, the same simulations as presented in this work can support derivation of such uncertainties.}. Once the errors have been added, the catalogue is cut with the selection criteria. For this analysis we apply cuts in CR in $[0.002,1]$~cts~s$^{-1}$ (roughly corresponding to $[0.28, 140] \times 10^{-14}$~ergs s$^{-1}$ cm$^{-2}$) and in HR in $[0.02 , 2.0]$.

\subsubsection{Cosmological analysis of mocks}
In order to recover the input cosmological parameters we employed a maximum likelihood method and sampled the cosmological parameters using a Markov Chain Monte Carlo (MCMC) method. For the description of the likelihood we made use of the unbinned Cash C-statistic \citep{Cash79} which provides a useful way of determining how well a given set of data fits the expected distribution. The log-likelihood which we compute for each set of cosmological parameters is given by,
\begin{eqnarray}
	\ln \mathcal{L} &=& \sum_{i} \ln \left(\frac{dn}{d\mathrm{CR}d\mathrm{HR}}(\mathrm{CR_i}, \mathrm{HR_i})\right) \nonumber \\
	&\ \ \ -& \int_{\mathrm{CR_{min}}}^{\mathrm{CR_{max}}}\int_{\mathrm{HR_{min}}}^{\mathrm{HR_{max}}}\frac{dn}{d\mathrm{CR}d\mathrm{HR}}d\mathrm{CR}d\mathrm{HR},
\end{eqnarray}
where the sum in the above equation runs over all selected clusters and the integral (calculated over the cluster selection criteria) gives the number of clusters expected to be within the CR-HR region. 

For this work, we chose to use the publicly available Python package \texttt{emcee} \citep{Foreman-Mackey2012a}, an affine invariant ensemble sampler.

For this analysis, we assume a $\Lambda$CDM cosmological model relying on the parameters calculated by \citet{hinshaw2013wmap}, in particular with $\Omega_\mathrm{m}=0.28$, $\Omega_\Lambda=0.72$, $\sigma_8=0.82$ and $H_0=70\ \mathrm{km\ s^{-1} Mpc^{-1}}$. The scaling relations for $M-T$ and $L-T$ are those derived by the XXL collaboration \citep{Pacaud2016,giles2016,lieu2016}. We only fit for two cosmological parameters, $\Omega_M$ and $\sigma_8$ since we only wish to show that incomplete knowledge of the selection function results in a bias to the recovered parameters. As shown in Fig.~\ref{fig:cludeteff} the eROSITA selection function is defined for a series of values for the core radius. Here we consider the effect of assuming a selection function which is defined only for a single value of 35~arcsec for the core radius. This core radius is obtained as a weighted average of the core radii (in arcminutes) of the X-CLASS sample of clusters  \cite{Clerc2012b, Ridl2017}. 

\begin{figure}
	\includegraphics[width=\columnwidth]{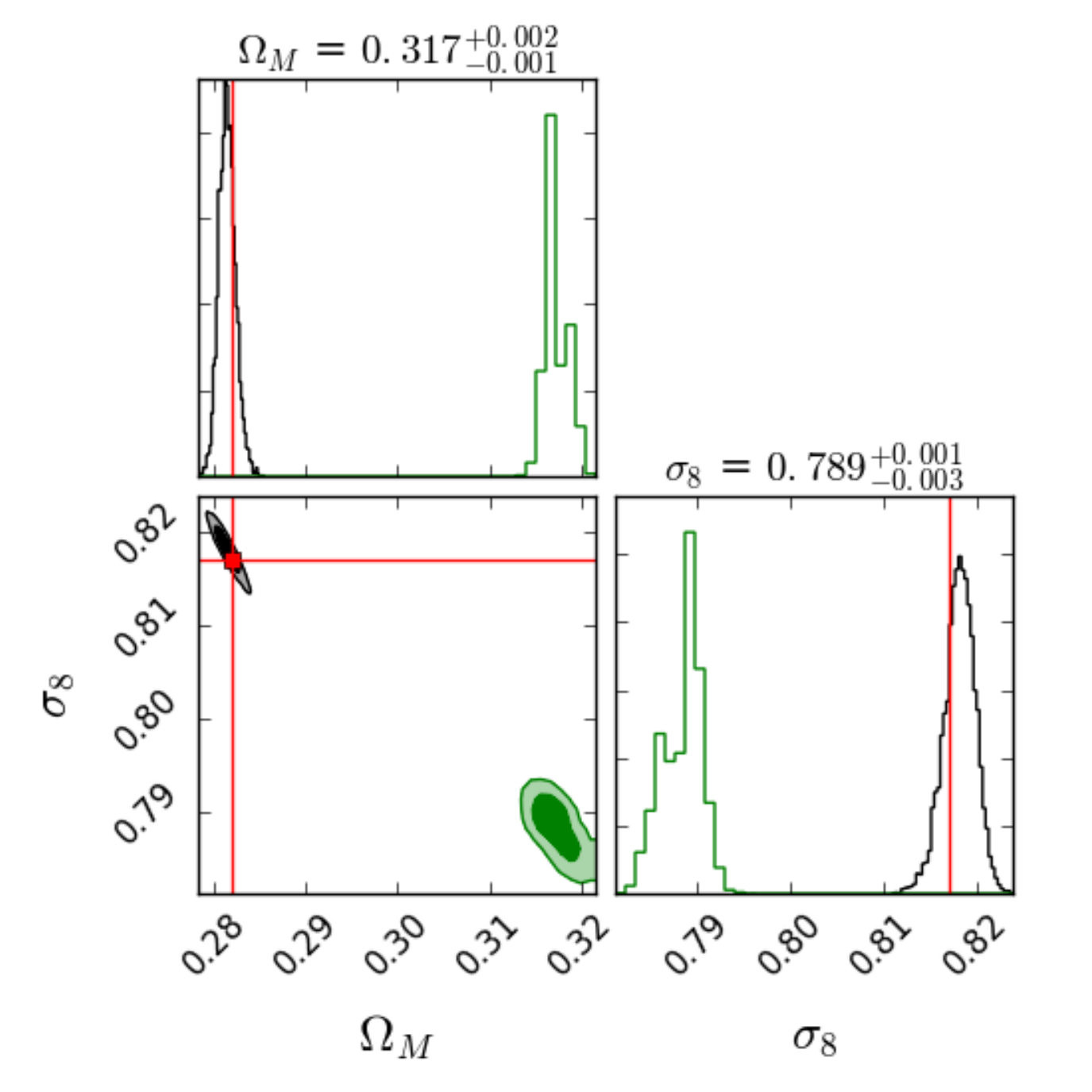}
      \caption{The bias introduced by the single core radius selection function. The black contours show the recovered constraints from the complete selection function while the green contours are the results obtained by fitting the cosmology assuming a single core radius in the selection function. The contours represent the 68\% and 95\% confidence intervals respectively. The red lines indicate the position of the fiducial input values used in the creation of the mock catalogue and the values quoted above the plots indicate the median value recovered when using the incorrect selection function.}
         \label{fig:cosmofit}
\end{figure}

A total of 104,574 clusters were generated over a hypothetical survey of 20,000 square degrees to a uniform depth of 1.6 ks. The selection criteria for clusters entering the mock were $ 0.002<\mathrm{CR}< 1.0$~cts\,s$^{-1}$ and $0.02 < \mathrm{HR} < 2.0$. The results obtained from the MCMC likelihood analysis are shown in Fig.~\ref{fig:cosmofit}. We see that very tight and unbiased constraints on both $\Omega_\mathrm{m}$ and $\sigma_8$ are obtained when the selection function is precisely known, as illustrated by the black contours in Fig.~\ref{fig:cosmofit}. On the other hand, a significant bias (shown by the green contours) is observed for both of these parameters when one assumes a core radius independent selection function when attempting to fit the cosmological parameters.

\section{\label{sect:simul_conclu}Conclusions}

We have produced and analyzed a set of realistic simulations for the eROSITA All-Sky Survey (eRASS) aiming towards precise selection functions for galaxy clusters. Our approach represents a trade-off between realism and tractability, capturing the essential (expected) instrumental and astrophysical features of the eRASS:
\begin{itemize}
\item fields of typical sizes in typical locations of the sky were selected and the exposure maps derived according to the spacecraft scanning law;
\item they are populated with active galactic nuclei following a realistic spectro-photometric distribution;
\item expected X-ray backgrounds (extragalactic and instrumental) are added;
\item the instrument is accurately modeled by using the \texttt{SIXTE} simulator, combined with accurate ray-tracing PSF and vignetting models as well as a detailed detector model;
\item galaxy clusters are simulated with various fluxes and sizes following an average $\beta$-model profile.
\end{itemize}

Our main result consists in a revisited selection function for extended sources defined in the (flux, extent) parameter space. We show that such a selection function can be coupled to cosmological codes and we provide an example with forward-modelling the entire galaxy cluster population with the CR-HR method \citep{Clerc2012}. Adjusting cosmological parameters to a mock catalog, we demonstrate that inaccurate knowledge of the selection function can lead to a significant bias in the derivation of cosmological parameters.

Such selection functions and results are valid to the extent of our current instrumental and astrophysical knowledge. Refined calibration and measurements (e.g.~background, point-spread function, etc.), on-ground and in-orbit, will provide updated results, critically needed for statistical analyses based on the eROSITA all-sky survey. Different source detection algorithms, possibly combining data from other wavelengths, may result in different quantitative selection functions; however the framework presented in this paper remains valid and it can be used to quickly and efficiently assess their ability to provide constraints on cosmological models of structure formation.


\begin{acknowledgements}
  
The authors thank the anonymous referee for its suggestions and comments which clearly increased the quality of this paper.
MERC acknowledges support by the German Aerospace Agency (DLR) with funds from the Ministry of Economy and Technology (BMWi) through grant 50 OR 1608. THR acknowledges support by the German Research Association (DFG) through grant RE 1462/6 and the Transregio 33 “The Dark Universe” sub-project B18. The authors thank S.~Grandis for useful comments on a preliminary version of this paper.

\end{acknowledgements}

\bibliographystyle{aa} 
\bibliography{myreferences} 

\begin{thebibliography}{55}
\expandafter\ifx\csname natexlab\endcsname\relax\def\natexlab#1{#1}\fi

\bibitem[{{Allen} {et~al.}(2011){Allen}, {Evrard}, \& {Mantz}}]{allen2011}
{Allen}, S.~W., {Evrard}, A.~E., \& {Mantz}, A.~B. 2011, \araa, 49, 409

\bibitem[{{Andreon} {et~al.}(2016){Andreon}, {Serra}, {Moretti}, \&
  {Trinchieri}}]{andreon2016}
{Andreon}, S., {Serra}, A.~L., {Moretti}, A., \& {Trinchieri}, G. 2016, \aap,
  585, A147

\bibitem[{{Arnaud}(1996)}]{arnaudxspec}
{Arnaud}, K.~A. 1996, in Astronomical Society of the Pacific Conference Series,
  Vol. 101, Astronomical Data Analysis Software and Systems V, ed. G.~H.
  {Jacoby} \& J.~{Barnes}, 17

\bibitem[{{B{\"o}hringer} {et~al.}(2017){B{\"o}hringer}, {Chon}, {Retzlaff},
  {Tr{\"u}mper}, {Meisenheimer}, \& {Schartel}}]{bohringer2017}
{B{\"o}hringer}, H., {Chon}, G., {Retzlaff}, J., {et~al.} 2017, \aj, 153, 220

\bibitem[{{B{\"o}hringer} {et~al.}(2004){B{\"o}hringer}, {Schuecker}, {Guzzo},
  {Collins}, {Voges}, {Cruddace}, {Ortiz-Gil}, {Chincarini}, {De Grandi},
  {Edge}, {MacGillivray}, {Neumann}, {Schindler}, \& {Shaver}}]{Boehringer2004}
{B{\"o}hringer}, H., {Schuecker}, P., {Guzzo}, L., {et~al.} 2004, \aap, 425,
  367

\bibitem[{{B{\"o}hringer} {et~al.}(2000){B{\"o}hringer}, {Voges}, {Huchra},
  {McLean}, {Giacconi}, {Rosati}, {Burg}, {Mader}, {Schuecker}, {Simi{\c c}},
  {Komossa}, {Reiprich}, {Retzlaff}, \& {Tr{\"u}mper}}]{bohringer2000}
{B{\"o}hringer}, H., {Voges}, W., {Huchra}, J.~P., {et~al.} 2000, \apjs, 129,
  435

\bibitem[{{B{\"o}hringer} \& {Werner}(2010)}]{BoehringerWerner2010}
{B{\"o}hringer}, H. \& {Werner}, N. 2010, \aapr, 18, 127

\bibitem[{{Borgani} {et~al.}(2001){Borgani}, {Rosati}, {Tozzi}, {Stanford},
  {Eisenhardt}, {Lidman}, {Holden}, {Della Ceca}, {Norman}, \&
  {Squires}}]{borgani2001}
{Borgani}, S., {Rosati}, P., {Tozzi}, P., {et~al.} 2001, \apj, 561, 13

\bibitem[{{Burenin} {et~al.}(2007){Burenin}, {Vikhlinin}, {Hornstrup},
  {Ebeling}, {Quintana}, \& {Mescheryakov}}]{burenin2007}
{Burenin}, R.~A., {Vikhlinin}, A., {Hornstrup}, A., {et~al.} 2007, \apjs, 172,
  561

\bibitem[{{Cash}(1979)}]{Cash79}
{Cash}, W. 1979, \apj, 228, 939

\bibitem[{{Cavaliere} \& {Fusco-Femiano}(1978)}]{cavaliere1978}
{Cavaliere}, A. \& {Fusco-Femiano}, R. 1978, \aap, 70, 677

\bibitem[{{Clerc} {et~al.}(2012){Clerc}, {Pierre}, {Pacaud}, \&
  {Sadibekova}}]{Clerc2012}
{Clerc}, N., {Pierre}, M., {Pacaud}, F., \& {Sadibekova}, T. 2012, \mnras, 423,
  3545

\bibitem[{Clerc {et~al.}(2012)Clerc, Sadibekova, Pierre, Pacaud, {Le Fevre},
  Adami, Altieri, \& Valtchanov}]{Clerc2012b}
Clerc, N., Sadibekova, T., Pierre, M., {et~al.} 2012, Mon. Not. R. Astron.
  Soc., 423, 3561

\bibitem[{{de Haan} {et~al.}(2016){de Haan}, {Benson}, {Bleem}, {Allen},
  {Applegate}, {Ashby}, {Bautz}, {Bayliss}, {Bocquet}, {Brodwin}, {Carlstrom},
  {Chang}, {Chiu}, {Cho}, {Clocchiatti}, {Crawford}, {Crites}, {Desai},
  {Dietrich}, {Dobbs}, {Doucouliagos}, {Foley}, {Forman}, {Garmire}, {George},
  {Gladders}, {Gonzalez}, {Gupta}, {Halverson}, {Hlavacek-Larrondo},
  {Hoekstra}, {Holder}, {Holzapfel}, {Hou}, {Hrubes}, {Huang}, {Jones},
  {Keisler}, {Knox}, {Lee}, {Leitch}, {von der Linden}, {Luong-Van}, {Mantz},
  {Marrone}, {McDonald}, {McMahon}, {Meyer}, {Mocanu}, {Mohr}, {Murray},
  {Padin}, {Pryke}, {Rapetti}, {Reichardt}, {Rest}, {Ruel}, {Ruhl},
  {Saliwanchik}, {Saro}, {Sayre}, {Schaffer}, {Schrabback}, {Shirokoff},
  {Song}, {Spieler}, {Stalder}, {Stanford}, {Staniszewski}, {Stark}, {Story},
  {Stubbs}, {Vanderlinde}, {Vieira}, {Vikhlinin}, {Williamson}, \&
  {Zenteno}}]{deHaan2016}
{de Haan}, T., {Benson}, B.~A., {Bleem}, L.~E., {et~al.} 2016, \apj, 832, 95

\bibitem[{{Ebeling} {et~al.}(2000){Ebeling}, {Edge}, {Allen}, {Crawford},
  {Fabian}, \& {Huchra}}]{ebeling2000}
{Ebeling}, H., {Edge}, A.~C., {Allen}, S.~W., {et~al.} 2000, \mnras, 318, 333

\bibitem[{{Foreman-Mackey} {et~al.}(2013){Foreman-Mackey}, {Hogg}, {Lang}, \&
  {Goodman}}]{Foreman-Mackey2012a}
{Foreman-Mackey}, D., {Hogg}, D.~W., {Lang}, D., \& {Goodman}, J. 2013, \pasp,
  125, 306

\bibitem[{{Forman} \& {Jones}(1982)}]{FormanJones1982}
{Forman}, W. \& {Jones}, C. 1982, \araa, 20, 547

\bibitem[{{Georgakakis} {et~al.}(2008){Georgakakis}, {Nandra}, {Laird}, {Aird},
  \& {Trichas}}]{georgakakis2008}
{Georgakakis}, A., {Nandra}, K., {Laird}, E.~S., {Aird}, J., \& {Trichas}, M.
  2008, \mnras, 388, 1205

\bibitem[{Giles {et~al.}(2016)Giles, Maughan, Pacaud, Lieu, Clerc, Pierre,
  Adami, Chiappetti, D{\'{e}}mocl{\'{e}}s, Ettori, {Le F{\'{e}}vre}, Ponman,
  Sadibekova, Smith, Willis, \& Ziparo}]{giles2016}
Giles, P.~A., Maughan, B.~J., Pacaud, F., {et~al.} 2016, Astron. Astrophys.,
  592, A3

\bibitem[{{Gilli} {et~al.}(1999){Gilli}, {Comastri}, {Brunetti}, \&
  {Setti}}]{gilli99}
{Gilli}, R., {Comastri}, A., {Brunetti}, G., \& {Setti}, G. 1999, \na, 4, 45

\bibitem[{{Gilli} {et~al.}(2007){Gilli}, {Comastri}, \& {Hasinger}}]{gilli07}
{Gilli}, R., {Comastri}, A., \& {Hasinger}, G. 2007, \aap, 463, 79

\bibitem[{{Giodini} {et~al.}(2013){Giodini}, {Lovisari}, {Pointecouteau},
  {Ettori}, {Reiprich}, \& {Hoekstra}}]{giodini2013}
{Giodini}, S., {Lovisari}, L., {Pointecouteau}, E., {et~al.} 2013, \ssr, 177,
  247

\bibitem[{{Gioia} {et~al.}(1990){Gioia}, {Henry}, {Maccacaro}, {Morris},
  {Stocke}, \& {Wolter}}]{Gioia1990}
{Gioia}, I.~M., {Henry}, J.~P., {Maccacaro}, T., {et~al.} 1990, \apjl, 356, L35

\bibitem[{{Hasinger} {et~al.}(2005){Hasinger}, {Miyaji}, \&
  {Schmidt}}]{Hasinger2005}
{Hasinger}, G., {Miyaji}, T., \& {Schmidt}, M. 2005, \aap, 441, 417

\bibitem[{{Hasselfield} {et~al.}(2013){Hasselfield}, {Hilton}, {Marriage},
  {Addison}, {Barrientos}, {Battaglia}, {Battistelli}, {Bond}, {Crichton},
  {Das}, {Devlin}, {Dicker}, {Dunkley}, {D{\"u}nner}, {Fowler}, {Gralla},
  {Hajian}, {Halpern}, {Hincks}, {Hlozek}, {Hughes}, {Infante}, {Irwin},
  {Kosowsky}, {Marsden}, {Menanteau}, {Moodley}, {Niemack}, {Nolta}, {Page},
  {Partridge}, {Reese}, {Schmitt}, {Sehgal}, {Sherwin}, {Sievers}, {Sif{\'o}n},
  {Spergel}, {Staggs}, {Swetz}, {Switzer}, {Thornton}, {Trac}, \&
  {Wollack}}]{hasselfield2013}
{Hasselfield}, M., {Hilton}, M., {Marriage}, T.~A., {et~al.} 2013, \jcap, 7,
  008

\bibitem[{{Henry} {et~al.}(2009){Henry}, {Evrard}, {Hoekstra}, {Babul}, \&
  {Mahdavi}}]{henry2009}
{Henry}, J.~P., {Evrard}, A.~E., {Hoekstra}, H., {Babul}, A., \& {Mahdavi}, A.
  2009, \apj, 691, 1307

\bibitem[{{Hinshaw} {et~al.}(2013){Hinshaw}, {Larson}, {Komatsu}, {Spergel},
  {Bennett}, {Dunkley}, {Nolta}, {Halpern}, {Hill}, {Odegard}, {Page}, {Smith},
  {Weiland}, {Gold}, {Jarosik}, {Kogut}, {Limon}, {Meyer}, {Tucker}, {Wollack},
  \& {Wright}}]{hinshaw2013wmap}
{Hinshaw}, G., {Larson}, D., {Komatsu}, E., {et~al.} 2013, \apjs, 208, 19

\bibitem[{{Klein} {et~al.}(2017){Klein}, {Mohr}, {Desai}, {Israel}, {Allam},
  {Benoit-L{\'e}vy}, {Brooks}, {Buckley-Geer}, {Carnero Rosell}, {Carrasco
  Kind}, {Cunha}, {da Costa}, {Dietrich}, {Eifler}, {Evrard}, {Frieman},
  {Gruen}, {Gruendl}, {Gutierrez}, {Honscheid}, {James}, {Kuehn}, {Lima},
  {Maia}, {March}, {Melchior}, {Menanteau}, {Miquel}, {Plazas}, {Reil},
  {Romer}, {Sanchez}, {Santiago}, {Scarpine}, {Schubnell}, {Sevilla-Noarbe},
  {Smith}, {Soares-Santos}, {Sobreira}, {Suchyta}, {Swanson}, {Tarle}, \& {the
  DES Collaboration}}]{klein2017}
{Klein}, M., {Mohr}, J.~J., {Desai}, S., {et~al.} 2017, ArXiv e-prints
  [\eprint[arXiv]{1706.06577}]

\bibitem[{{Kolodzig} {et~al.}(2017){Kolodzig}, {Gilfanov}, {H{\"u}tsi}, \&
  {Sunyaev}}]{kolodzig2017}
{Kolodzig}, A., {Gilfanov}, M., {H{\"u}tsi}, G., \& {Sunyaev}, R. 2017, \mnras,
  466, 3035

\bibitem[{{Lehmer} {et~al.}(2012){Lehmer}, {Xue}, {Brandt}, {Alexander},
  {Bauer}, {Brusa}, {Comastri}, {Gilli}, {Hornschemeier}, {Luo}, {Paolillo},
  {Ptak}, {Shemmer}, {Schneider}, {Tozzi}, \& {Vignali}}]{lehmer2012}
{Lehmer}, B.~D., {Xue}, Y.~Q., {Brandt}, W.~N., {et~al.} 2012, \apj, 752, 46

\bibitem[{Lieu {et~al.}(2016)Lieu, Smith, Giles, Ziparo, Maughan,
  D{\'{e}}mocl{\`{e}}s, Pacaud, Pierre, Adami, Bah{\'{e}}, Clerc, Chiappetti,
  Eckert, Ettori, Lavoie, {Le Fevre}, McCarthy, Kilbinger, Ponman, Sadibekova,
  \& Willis}]{lieu2016}
Lieu, M., Smith, G.~P., Giles, P.~A., {et~al.} 2016, Astron. Astrophys., 592,
  A4

\bibitem[{{Lovisari} {et~al.}(2015){Lovisari}, {Reiprich}, \&
  {Schellenberger}}]{lovisari2015}
{Lovisari}, L., {Reiprich}, T.~H., \& {Schellenberger}, G. 2015, \aap, 573,
  A118

\bibitem[{{Lumb} {et~al.}(2002){Lumb}, {Warwick}, {Page}, \& {De
  Luca}}]{lumb2002}
{Lumb}, D.~H., {Warwick}, R.~S., {Page}, M., \& {De Luca}, A. 2002, \aap, 389,
  93

\bibitem[{{Mantz} {et~al.}(2010{\natexlab{a}}){Mantz}, {Allen}, {Ebeling},
  {Rapetti}, \& {Drlica-Wagner}}]{mantz2010b}
{Mantz}, A., {Allen}, S.~W., {Ebeling}, H., {Rapetti}, D., \& {Drlica-Wagner},
  A. 2010{\natexlab{a}}, \mnras, 406, 1773

\bibitem[{{Mantz} {et~al.}(2010{\natexlab{b}}){Mantz}, {Allen}, {Rapetti}, \&
  {Ebeling}}]{mantz2010}
{Mantz}, A., {Allen}, S.~W., {Rapetti}, D., \& {Ebeling}, H.
  2010{\natexlab{b}}, \mnras, 406, 1759

\bibitem[{{Mantz} {et~al.}(2014){Mantz}, {Allen}, {Morris}, {Rapetti},
  {Applegate}, {Kelly}, {von der Linden}, \& {Schmidt}}]{mantz2014}
{Mantz}, A.~B., {Allen}, S.~W., {Morris}, R.~G., {et~al.} 2014, \mnras, 440,
  2077

\bibitem[{{Merloni} {et~al.}(2012){Merloni}, {Predehl}, {Becker},
  {B{\"o}hringer}, {Boller}, {Brunner}, {Brusa}, {Dennerl}, {Freyberg},
  {Friedrich}, {Georgakakis}, {Haberl}, {Hasinger}, {Meidinger}, {Mohr},
  {Nandra}, {Rau}, {Reiprich}, {Robrade}, {Salvato}, {Santangelo}, {Sasaki},
  {Schwope}, {Wilms}, \& {German eROSITA Consortium}}]{merloni2012}
{Merloni}, A., {Predehl}, P., {Becker}, W., {et~al.} 2012, ArXiv e-prints
  [\eprint[arXiv]{1209.3114}]

\bibitem[{{Nurgaliev} {et~al.}(2017){Nurgaliev}, {McDonald}, {Benson}, {Bleem},
  {Bocquet}, {Forman}, {Garmire}, {Gupta}, {Hlavacek-Larrondo}, {Mohr},
  {Nagai}, {Rapetti}, {Stark}, {Stubbs}, \& {Vikhlinin}}]{nurgaliev2017}
{Nurgaliev}, D., {McDonald}, M., {Benson}, B.~A., {et~al.} 2017, \apj, 841, 5

\bibitem[{{Pacaud} {et~al.}(2016){Pacaud}, {Clerc}, {Giles}, {Adami},
  {Sadibekova}, {Pierre}, {Maughan}, {Lieu}, {Le F{\`e}vre}, {Alis}, {Altieri},
  {Ardila}, {Baldry}, {Benoist}, {Birkinshaw}, {Chiappetti},
  {D{\'e}mocl{\`e}s}, {Eckert}, {Evrard}, {Faccioli}, {Gastaldello}, {Guennou},
  {Horellou}, {Iovino}, {Koulouridis}, {Le Brun}, {Lidman}, {Liske},
  {Maurogordato}, {Menanteau}, {Owers}, {Poggianti}, {Pomar{\`e}de}, {Pompei},
  {Ponman}, {Rapetti}, {Reiprich}, {Smith}, {Tuffs}, {Valageas}, {Valtchanov},
  {Willis}, \& {Ziparo}}]{Pacaud2016}
{Pacaud}, F., {Clerc}, N., {Giles}, P.~A., {et~al.} 2016, \aap, 592, A2

\bibitem[{{Pacaud} {et~al.}(2007){Pacaud}, {Pierre}, {Adami}, {Altieri},
  {Andreon}, {Chiappetti}, {Detal}, {Duc}, {Galaz}, {Gueguen}, {Le F{\`e}vre},
  {Hertling}, {Libbrecht}, {Melin}, {Ponman}, {Quintana}, {Refregier},
  {Sprimont}, {Surdej}, {Valtchanov}, {Willis}, {Alloin}, {Birkinshaw},
  {Bremer}, {Garcet}, {Jean}, {Jones}, {Le F{\`e}vre}, {Maccagni}, {Mazure},
  {Proust}, {R{\"o}ttgering}, \& {Trinchieri}}]{Pacaud2007}
{Pacaud}, F., {Pierre}, M., {Adami}, C., {et~al.} 2007, \mnras, 382, 1289

\bibitem[{{Pacaud} {et~al.}(2006){Pacaud}, {Pierre}, {Refregier}, {Gueguen},
  {Starck}, {Valtchanov}, {Read}, {Altieri}, {Chiappetti}, {Gandhi}, {Garcet},
  {Gosset}, {Ponman}, \& {Surdej}}]{Pacaud2006}
{Pacaud}, F., {Pierre}, M., {Refregier}, A., {et~al.} 2006, \mnras, 372, 578

\bibitem[{{Pillepich} {et~al.}(2012){Pillepich}, {Porciani}, \&
  {Reiprich}}]{Pillepich2012}
{Pillepich}, A., {Porciani}, C., \& {Reiprich}, T.~H. 2012, \mnras, 422, 44

\bibitem[{{Planck Collaboration} {et~al.}(2016){Planck Collaboration}, {Ade},
  {Aghanim}, {Arnaud}, {Ashdown}, {Aumont}, {Baccigalupi}, {Banday},
  {Barreiro}, {Bartlett}, \& et~al.}]{planck2016}
{Planck Collaboration}, {Ade}, P.~A.~R., {Aghanim}, N., {et~al.} 2016, \aap,
  594, A24

\bibitem[{{Predehl}(2017)}]{predehl2017}
{Predehl}, P. 2017, Astronomische Nachrichten, 338, 159

\bibitem[{{Ragagnin} {et~al.}(2016){Ragagnin}, {Dolag}, {Biffi}, {Cadolle Bel},
  {Hammer}, {Krukau}, {Petkova}, \& {Steinborn}}]{dolag2016}
{Ragagnin}, A., {Dolag}, K., {Biffi}, V., {et~al.} 2016, ArXiv e-prints
  [\eprint[arXiv]{1612.06380}]

\bibitem[{Ridl {et~al.}(2017)Ridl, Clerc, Sadibekova, Faccioli, Pacaud,
  Greiner, Kr{\"u}hler, Rau, Salvato, Menzel, {et~al.}}]{Ridl2017}
Ridl, J., Clerc, N., Sadibekova, T., {et~al.} 2017, Monthly Notices of the
  Royal Astronomical Society, 468, 662

\bibitem[{{Rozo} {et~al.}(2014){Rozo}, {Bartlett}, {Evrard}, \&
  {Rykoff}}]{rozo2014}
{Rozo}, E., {Bartlett}, J.~G., {Evrard}, A.~E., \& {Rykoff}, E.~S. 2014,
  \mnras, 438, 78

\bibitem[{{Sadibekova} {et~al.}(2014){Sadibekova}, {Pierre}, {Clerc},
  {Faccioli}, {Gastaud}, {Le Fevre}, {Rozo}, \& {Rykoff}}]{sadibekova2014}
{Sadibekova}, T., {Pierre}, M., {Clerc}, N., {et~al.} 2014, \aap, 571, A87

\bibitem[{Schmid(2012)}]{schmid2012}
Schmid, C. 2012, PhD thesis, Uni. Erlangen

\bibitem[{{Stanek} {et~al.}(2006){Stanek}, {Evrard}, {B{\"o}hringer},
  {Schuecker}, \& {Nord}}]{stanek2006}
{Stanek}, R., {Evrard}, A.~E., {B{\"o}hringer}, H., {Schuecker}, P., \& {Nord},
  B. 2006, \apj, 648, 956

\bibitem[{{Tenzer} {et~al.}(2010){Tenzer}, {Warth}, {Kendziorra}, \&
  {Santangelo}}]{tenzer2010}
{Tenzer}, C., {Warth}, G., {Kendziorra}, E., \& {Santangelo}, A. 2010, in
  \procspie, Vol. 7742, High Energy, Optical, and Infrared Detectors for
  Astronomy IV, 77420Y

\bibitem[{{Vikhlinin} {et~al.}(2009){Vikhlinin}, {Kravtsov}, {Burenin},
  {Ebeling}, {Forman}, {Hornstrup}, {Jones}, {Murray}, {Nagai}, {Quintana}, \&
  {Voevodkin}}]{vikhlinin2009}
{Vikhlinin}, A., {Kravtsov}, A.~V., {Burenin}, R.~A., {et~al.} 2009, \apj, 692,
  1060

\bibitem[{{Vikhlinin} {et~al.}(1998){Vikhlinin}, {McNamara}, {Forman}, {Jones},
  {Quintana}, \& {Hornstrup}}]{vikhlinin1998}
{Vikhlinin}, A., {McNamara}, B.~R., {Forman}, W., {et~al.} 1998, \apj, 502, 558

\bibitem[{{Wen} {et~al.}(2012){Wen}, {Han}, \& {Liu}}]{wen2012}
{Wen}, Z.~L., {Han}, J.~L., \& {Liu}, F.~S. 2012, \apjs, 199, 34

\bibitem[{{Yaqoob} {et~al.}(1997){Yaqoob}, {McKernan}, {Ptak}, {Nandra}, \&
  {Serlemitsos}}]{Yaqoob1997}
{Yaqoob}, T., {McKernan}, B., {Ptak}, A., {Nandra}, K., \& {Serlemitsos}, P.~J.
  1997, \apjl, 490, L25

\end{thebibliography}

\appendix
\section{Analytic fit to the point-like and extended source selection curves}
    \label{sect:modelfits}

We provide analytic functions that represent the results obtained in Figures~\ref{fig:agnsel} and~\ref{fig:cludeteff}.
Due to the limited number of points sampling the curves in the steep transition region, we fitted functions that constitute a reasonable representation of the simulation.

 For the extended source selection (galaxy clusters), we parametrize the completeness, dubbed $c$, as a function of 0.5 -- 2~keV flux, exposure time ($T_{exp}$) and core radius ($r_c$) as follows :

 \begin{align*}
 a(T,R) & = 13.5-(R-1.2)^2 + (T-3.204)/1.28 \\
 c(F, T, R) & = 0.5+0.5\, \textrm{erf} \left( ( F + a(T, R)) / 0.2 \right),
 \end{align*}

where erf represents the error function, 
\begin{align*}
T & =\log_{10}(T_\textrm{exp}/[{\rm ks}])\\
R & =\log_{10}(r_\textrm{c}/[{\rm arcsec}])\\
F & =\log_{10}({\rm Flux}/[{\rm erg\, cm}^{-2} {\, \rm s}^{-1}]). 
\end{align*}
For the point-like sources, the parametrization only depends on flux and exposure time:

\begin{equation}
c (F, T) = 0.5 + 0.5 \, {\rm erf}
\left(
\frac{F+0.5936(T-3.1828)}{0.3204}
\right)
\end{equation}

\begin{figure}
	\includegraphics[width=\hsize,trim=0 0 0 1.2cm,clip]{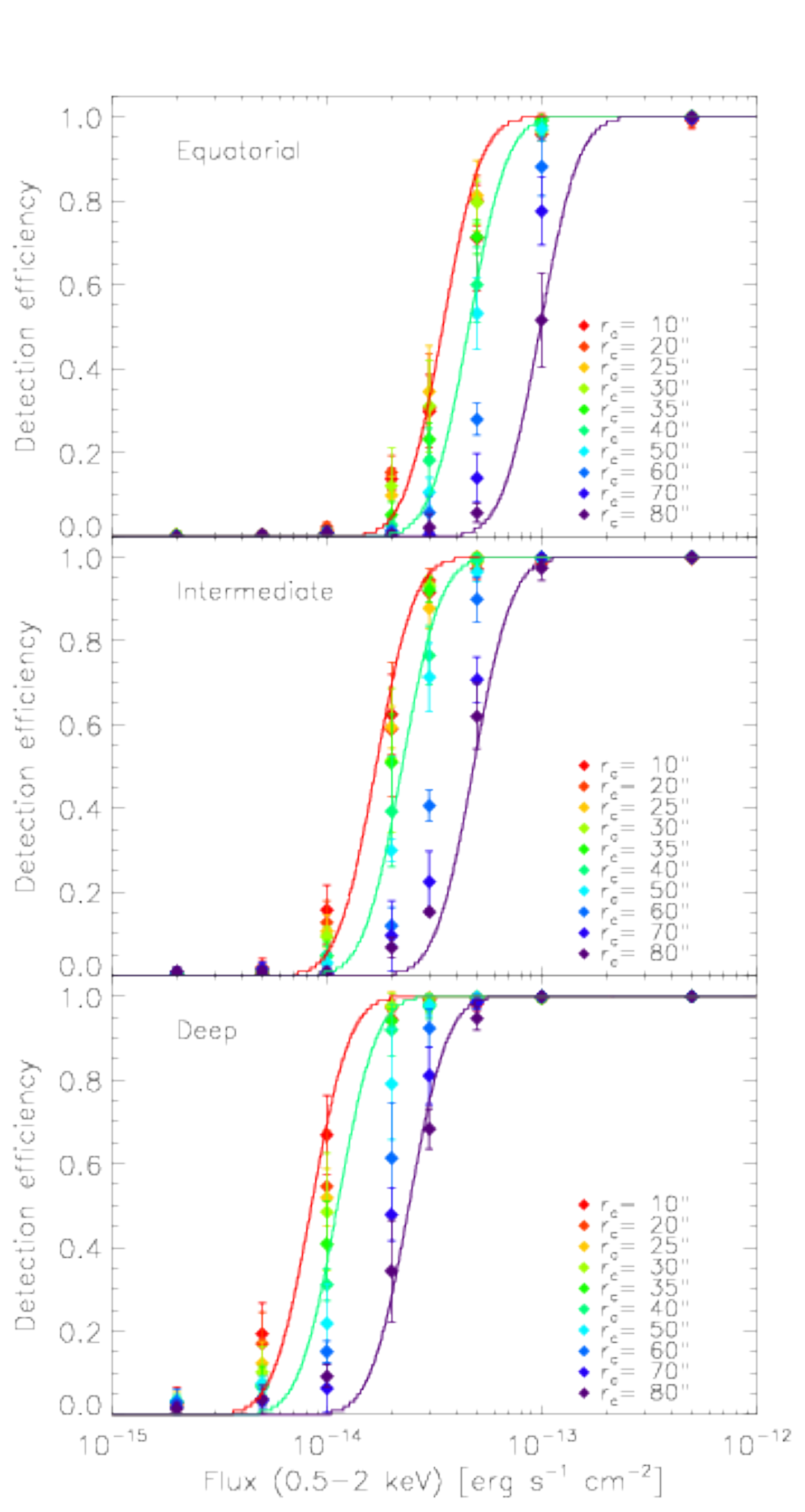}
      \caption{Figure similar to Fig.~\ref{fig:cludeteff}, where we superimposed the model lines computed according to formulas in App.~\ref{sect:modelfits}, for $r_c=10,~40,~80\arcsec$.}
         \label{fig:cludeteff_fit}
\end{figure}

In figures \ref{fig:agnsel} and~\ref{fig:cludeteff_fit} we show the models and their relatively good agreement to the data points extracted from the simulations. Such simple models cannot fully account for the details of the selection function curves, but they should be useful to provide ready-to-use estimates of completeness for various forecasts.

\end{document}